\title{Efficient Algorithms for Searching Optimal Shortened Cyclic
Single-Burst-Correcting Codes}
\author{Luis Javier Garc\'{\i}a Villalba$^1$,
Jos\'e Ren\'e Fuentes Cortez$^1$, \\Ana Lucila Sandoval Orozco$^1$
and Mario Blaum $^{1,2}{\ }$\footnote{This work was supported by the
Ministerio de Ciencia e Innovaci\'on (MICINN, Spain) through Project
TEC2010-18894/TCM and the Ministerio de Industria, Turismo y Comercio
(MITyC, Spain) through Project AVANZA COMPETITIVIDAD I+D+I TSI-020100-2010-482.}\\
\\$^1$ Grupo de An\'{a}lisis, Seguridad y Sistemas (GASS)\\
Facultad de Inform\'atica, Despacho 431\\
Universidad Complutense de Madrid (UCM)\\
C/ Profesor Jos\'e Garc\'{\i}a Santesmases s/n, 28040 Madrid, Spain\\
{\tt \{javiergv,jrfuente,asandoval,mario.blaum\}@fdi.ucm.es}\\
\\$^2$ IBM Almaden Research Center\\
650 Harry Road\\
San Jose, CA 95120, USA\\
{\tt mblaum@us.ibm.com}
}
\documentclass[12pt]{article}
\usepackage{latexsym}
 \newtheorem{lemma}{Lemma}[section]
 
 \newtheorem{defin}{Definition}[section]
 \newtheorem{ex}{Example}[section]
 \newtheorem{alg}{Algorithm}[section]

\newtheorem{COROLLARY}{\indent Corollary}

\newtheorem{EXAMPLE}{\indent Example}

\newtheorem{THEOREM}{\indent Theorem}

\newtheorem{REMARK}{\indent Remark}

\newcommand{\fullstop}{\hspace{-0.85em} {\bf .}}

\newcommand{\uh}{\mbox{$\underline{h}$}}
\newcommand{\uv}{\mbox{$\underline{v}$}}

\newcommand{\us}{\mbox{$\underline{s}$}}

\newcommand{\uc}{\mbox{$\underline{c}$}}

\newcommand{\ug}{\mbox{$\underline{g}$}}

\newcommand{\la}{\mbox{$\leftarrow$}}

\newcommand{\eq}{\mbox{$\, =\,$}}

\newcommand{\lan}{\mbox{$\langle$}}
\newcommand{\ran}{\mbox{$\rangle$}}

\newcommand{\qed}{\hfill$\Box$\\[1ex]}

\newcommand{\uw}{\mbox{$\underline{w}$}}

\newcommand{\xor}{\mbox{$\oplus$}}

\newcommand{\C}{\mbox{${\cal C}$}}

\newcommand{\G}{\mbox{${\cal G}$}}

\newcommand{\br}{\\ }
\newcommand{\ce}{\begin{center}}
\newcommand{\cen}{\end{center}}

\newcommand{\ipb}{\begin{description}}
\newcommand{\ipn}{\end{description}}

\newcommand{\qb}{\begin{quote}}
\newcommand{\qn}{\end{quote}}

\newcommand{\tp}{\begin{titlepage}}
\newcommand{\tpn}{\end{titlepage}}
\newcommand{\zb}{\begin{figure}[hbtp]}
\newcommand{\zn}{\end{figure}}

\newcommand{\EQX}[1]{\begin{equation}\label{#1}}
\newcommand{\ENX}{\end{equation}}
\newcommand{\EQL}{\begin{eqnarray*}}
\newcommand{\ENL}{\end{eqnarray*}}
\newcommand{\EQLX}[1]{\begin{eqnarray}\label{#1}}
\newcommand{\ENLX}{\end{eqnarray}}

\newcommand{\open}{\begin{document}}
\newcommand{\close}{\end{document}}
\newcommand{\lfcr}[1]{\br\hspace*{#1em}}
\newenvironment{mat}[1]
{\left[\begin{array}{#1}}{\end{array}\right]}
\newcommand{\GAMMA}{\Gamma}
\newcommand{\DELTA}{\Delta}
\newcommand{\THETA}{\Theta}
\newcommand{\LAMBDA}{\Lambda}
\newcommand{\XI}{\Xi}
\newcommand{\PI}{\Pi}
\newcommand{\SIGMA}{\Sigma}
\newcommand{\UPSILON}{\Upsilon}
\newcommand{\PHI}{\Phi}
\newcommand{\PSI}{\Psi}
\newcommand{\OMEGA}{\Omega}
\newcommand{\bldgreek}[1]{\mbox{\boldmath $#1$}}
\newcommand{\bldbeta}{\bldgreek{\beta}}
\newcommand{\bldgamma}{\bldgreek{\gamma}}
\newcommand{\blddelta}{\bldgreek{\delta}}
\newcommand{\bldepsilon}{\bldgreek{\epsilon}}
\newcommand{\bldvarepsilon}{\bldgreek{\varepsilon}}
\newcommand{\bldzeta}{\bldgreek{\zeta}}
\newcommand{\bldeta}{\bldgreek{\eta}}
\newcommand{\bldtheta}{\bldgreek{\theta}}
\newcommand{\bldvartheta}{\bldgreek{\vartheta}}
\newcommand{\bldiota}{\bldgreek{\iota}}
\newcommand{\bldkappa}{\bldgreek{\kappa}}
\newcommand{\bldlambda}{\bldgreek{\lambda}}
\newcommand{\bldmu}{\bldgreek{\mu}}
\newcommand{\bldnu}{\bldgreek{\nu}}
\newcommand{\bldxi}{\bldgreek{\xi}}
\newcommand{\bldpi}{\bldgreek{\pi}}
\newcommand{\bldvarpi}{\bldgreek{\varpi}}
\newcommand{\bldrho}{\bldgreek{\rho}}
\newcommand{\bldvarrho}{\bldgreek{\varrho}}
\newcommand{\bldsigma}{\bldgreek{\sigma}}
\newcommand{\bldvarsigma}{\bldgreek{\varsigma}}
\newcommand{\bldtau}{\bldgreek{\tau}}
\newcommand{\bldupsilon}{\bldgreek{\upsilon}}
\newcommand{\bldphi}{\bldgreek{\phi}}
\newcommand{\bldvarphi}{\bldgreek{\varphi}}
\newcommand{\bldchi}{\bldgreek{\chi}}
\newcommand{\bldpsi}{\bldgreek{\psi}}
\newcommand{\bldomega}{\bldgreek{\omega}}
\begin{document}
\parindent=10pt
\maketitle
\begin{abstract}
In a previous work it was shown that the best measure for
the efficiency of a single burst-correcting code is obtained using the
Gallager bound as opposed to the Reiger bound. In
this paper, an
efficient algorithm that searches for the best (shortened)
cyclic burst-correcting codes is presented. Using this algorithm,
extensive tables that either tie existing constructions or improve
them are obtained for burst lengths up to $b\eq 10$.
\end{abstract}



\section{Introduction}
\label{intro}
In~\cite{gfb}, we studied the efficiency of linear burst-correcting
block codes and we argued that the framework for determining such
efficiency is based not on the commonly used Reiger bound~\cite{r} but on the
Gallager bound~\cite{g}. Using this new framework, in particular we provided
tables of shortened cyclic single burst-correcting codes with
efficiencies either equal or larger than the ones of existing codes.
Such new codes were obtained
by computer search. However, absent from the
study in~\cite{gfb}, is the search algorithm utilized in order to
obtain the codes. In this paper, we present such search algorithm,
which allows for fast searches given a variety of parameters.
Actually, some of the best single burst-correcting codes have been
found by computer search~\cite{kasami}\cite{mm}, usually improving
the parameters of the best known family of single burst-correcting
cyclic codes, the Fire codes~\cite{f}. Many of the results of such
searches can be found in the tables given in~\cite{lc}\cite{lc2}.
However, in this paper we search for codes by taking into account the
efficiency criterion developed in~\cite{gfb}.

In order to make the paper self-contained, we repeat several of the
definitions and concepts of~\cite{gfb} but with less level of detail.
Let us start with the definition of a burst with respect to
a guard space~\cite{lc}\cite{lc2}:

\begin{defin}
\label{guard}
{\em
Assume that an all-zero sequence is transmitted and let $e_0,e_1,e_2\ldots$
be the received sequence,
i.e., 1s represent errors and 0s absence of errors. Then, a vector of
$b$ consecutive bits $(e_l,e_{l+1},\ldots,e_{l+b-1})$ is called a burst
of length $b$ with respect to a guard space of length $g$ if:

\begin{enumerate}

\item $e_l\eq e_{l+b-1}\eq 1$.

\item $b\leq g$.

\item The $g$ bits preceding $e_l$ and the $g$ bits following
$e_{l+b-1}$ are all 0s (if $l<g$ then all the bits preceding $l$
are 0).\qed

\end{enumerate}
}
\end{defin}

Assume that we encode a (semi-infinite) sequence using a code $\C$
(either block or convolutional). If a block code of length $n$ is
used, the encoded sequence is divided into blocks of length $n$.
So let us define the burst-correcting capability of code $\C$.

\begin{defin}
\label{bc}
{\em
Assume that an encoded sequence under code $\C$ is transmitted into a
channel, a (possibly
noisy) version is received and the non-zero elements (i.e.,
the bits in error) in the difference between the
received sequence and the transmitted sequence can be grouped in
bursts of
length at most $b$ with guard space $g$. If code $\C$ can correct any
such received sequence, we say that $\C$ is a
$(b,g)$-burst-correcting code.\qed
}
\end{defin}

The values required for the pair $(b,g)$ can sometimes be determined from
the statistics of the channel. For instance, a well known model for
isolated bursts is given by the Gilbert-Elliot
channel~\cite{gi}\cite{el}.

Assume that the pair $(b,g)$ is given and that we
want to construct a $(b,g)$-burst-correcting code with rate as large as
possible. We will search only for either cyclic or shortened cyclic block
codes.

Next, we consider single-burst-correcting $[n,k]$ linear binary
codes and we will see how they relate to our $(b,g)$-burst-correcting
model. When we say that an $[n,k]$ code $\C$ can correct a single
burst of length up to $b$, there are two types of bursts: non-all
around (NAA) and all-around (AA) bursts. Let us
define them formally.

\begin{defin}
\label{AA}
{\em
Given a block of $n$ bits $e_0,e_1,\ldots,e_{n-1}$, we say that
$e_l,e_{l+1},\ldots, e_{l+b-1}$ is a NAA burst of length $b$ for
$0\leq l\leq n-1$, if
$l+b\leq n$, $e_l\eq e_{l+b-1}\eq 1$ and
$e_i\eq 0$ for $i<l$ and $i>l+b-1$.

Similarly, given a block of $n$ bits $e_0,e_1,\ldots,e_{n-1}$, we say that\\
$e_l,e_{l+1},\ldots, e_{n-1},e_0,e_1,\ldots ,e_{l+b-n-1}$
is an AA burst of length $b$, where $1\leq l\leq n-1$ and $b<n$,
if $l+b> n$, $e_l\eq e_{l+b-n-1}\eq 1$ and
$e_i\eq 0$ for $l+b-n-1<i<l$.\qed
}
\end{defin}

AA bursts have received different names in literature. In~\cite{b},
bursts of this type are called {\em cyclic} (a name we prefer to
avoid in order to prevent confusion with cyclic codes).
In~\cite{mm}, NAA bursts are called {\em open-loop} bursts and AA
bursts are called {\em closed-loop} bursts, while in~\cite{ht}, AA
bursts are called {\em wrap-around} bursts.

\begin{defin}
\label{d1}
{\em
Consider an $[n,k]$ code $\C$. If $\C$ can correct up to
a single NAA burst of length up to $b$, or up to
a single AA burst of length
up to $\ell$, then we say that $\C$ is an $[n,k,\lan b,\ell\ran]$
burst-correcting code. \qed
}
\end{defin}

The following lemma~\cite{gfb} is immediate and it connects
Definitions~\ref{bc} and~\ref{d1}:

\begin{lemma}
\label{lguardl}
{\em
Let $\C$ be an $[n,k,\lan b,\ell\ran]$ burst-correcting code,
$1\leq\ell\leq b$.
Then $\C$ is a
$(b,n-\ell)$ burst-correcting code.
}
\end{lemma}

The following lemma is also immediate from Definition~\ref{d1} and
Lemma~\ref{lguardl}:

\begin{lemma}
\label{l2}
{\em
Let $\C$ be an $[n,k,\lan b,\ell\ran]$ burst-correcting code,
$2\leq\ell\leq b$.
Then $\C$ is an $[n,k,\lan b,{\ell -1}\ran]$
$(b,n-\ell+1)$-burst-correcting code.
}
\end{lemma}

The following lemma is simple and well known (see for
instance~\cite{mm}), but let us put it in
the framework of Definition~\ref{d1}:

\begin{lemma}
\label{l3}
{\em
Assume that $\C$ is an $[n,k,\lan b,1\ran]$ burst-correcting {\em cyclic }code.
Then $\C$ is an $[n,k,\lan b,b\ran]$ burst-correcting code.
}
\end{lemma}

The best single-burst-correcting codes considered in literature prior
to~\cite{gfb} are either
$\lan b,1\ran$ or $\lan b,b\ran$ single-burst-correcting codes~\cite{lc}\cite{lc2}.
It was shown in~\cite{gfb} that by taking intermediate values $1<\ell
<b$, codes with better rates were often found.

Given $(b,g)$, we will proceed as follows:
for each $\ell$, $1\leq \ell\leq b$, we search for an optimal
$[g+\ell,k_{\ell},\lan b,\ell\ran]$
burst-correcting code (that by Lemma~\ref{lguardl} is
$(b,g)$-burst-correcting) by using the search algorithm to be described
in the next section. We have denoted the dimension of each
code by $k_{\ell}$ to indicate its dependance on $\ell$.
Then we
choose the code that gives us the
largest value of the rate $k_{\ell}/(g+\ell)$ (or, in other words, the one
that maximizes the Gallager efficiency~\cite{gfb}). The next example,
taken from~\cite{gfb}, shows that given $(b,g)$, sometimes there are
values of $\ell$ that are
neither 1 nor $b$ but that give codes with better rates than the
former:

\begin{ex}
\label{exint}
{\rm Consider a pair $(b,g)\eq (3,25)$.
There is a shortened cyclic $[28,19,\lan 3,3\ran]$ (3,25)-burst-correcting code
generated by $x^9+x^8+x^6+1$. By computer
search we can determine that there is no $[28,20,\lan 3,3\ran]$
burst-correcting (shortened) cyclic code.

Similarly, we find that there are $[26,19,\lan 3,1\ran]$
(3,25)-burst-correcting shortened cyclic
codes, but not $[26,20,\lan 3,1\ran]$ codes.

However, the $[27,20,\lan 3,2\ran]$ shortened cyclic code generated
by $x^7+x^6+x^3+1$
is a (3,25)-burst-correcting code and it has better rate than both
the $[28,19,\lan 3,3\ran]$ and the $[26,19,\lan 3,1\ran]$ codes. \qed
}
\end{ex}

In the next section, we present the search algorithm for shortened
cyclic codes that are $(b,g)$ burst-correcting. Let us point out that
the search algorithm presented in~\cite{mm}
finds the burst-error capability of a given {\em cyclic} code.
The search algorithm presented in~\cite{kasami} does not take into
account the guard space explicitly.

\section{Algorithm searching for optimal burst-correcting codes}
\label{search}


In order to check if there exists
an $[n,k,\lan b,\ell\ran]$ (shortened) cyclic code, $1\leq \ell\leq b$,
we need to check all possible generator polynomials of degree $n-k$.
If we find one, we
stop the search. If there is none, then we try to find an $[n,k-1]$
code using the same procedure, and so on, until we determine the
largest possible value of $k$.

Many polynomials can be eliminated from the search with a quick test.
A generator polynomial
$g(x)$ may be represented as a binary vector.
We may assume without loss of generality that such binary vector
begins and ends with a 1. Moreover, it can be proven without much
difficulty that $g(x)$ generates an $[n,k,\lan b,\ell\ran]$
(shortened) cyclic code, if
and only if the code generated by the polynomial obtained by reversing
the order of the bits of $g(x)$ is also an $[n,k,\lan b,\ell\ran]$
(shortened) cyclic code.
This observation allows to simplify the search: if we have
found out that the code generated by
$g(x)$ is not an $[n,k,\lan b,\ell\ran]$ code,
then it is not necessary to test the code generated by $g(x)$ in
reverse order.

Another simple test when checking if the code
generated by $g(x)$ is an $[n,k,\lan b,\ell\ran]$ code, is to measure
the burst-$b$ weight~\cite{ww} of $g(x)$. If such
burst-$b$ weight is
smaller than 3, this means that in particular the code cannot be
an $[n,k,\lan b,\ell\ran]$ code, so no further tests on $g(x)$ are
necessary
and we may proceed with the next candidate polynomial.
It is easy to determine all the generator polynomials of burst-$b$ weight
smaller than 3, if we take into account that $g(x)$
must start and end with a 1. Writing $g(x)$ as a vector $\ug$ of
length $n-k+1$, we have $\ug\eq (g_0,g_1,\ldots,g_{n-k})$, where
$g_0\eq g_{n-k}\eq 1$. If $\ug$ has burst-$b$ weight
smaller than 3, it means that its non-zero entries can
be covered by at most two bursts of length up to $b$ each. But there
are exactly $2^{2b-2}$ vectors $\ug$ that have
burst-$b$ weight smaller than 3: they are all the vectors
$\ug\eq (g_0,g_1,\ldots,g_{n-k})$ with $g_0\eq g_{n-k}\eq 1$ such that
$g_i\eq 0$ for $b\leq i\leq n-k-b$. So these $2^{2b-2}$ vectors can
also be eliminated from the search.

For example,
assume that we want to test the polynomial $g(x)\eq 1+x^3+x^8$ and
see if it
generates a $[12,4,\lan 3,\ell\ran]$ code, $1\leq\ell\leq 3$. Writing
$g(x)$ as a vector of length 9 $\ug$, we have

\begin{eqnarray*}
\ug &=&(1\;0\;0\;1\;0\;0\;0\;0\;1)
\end{eqnarray*}

Notice that $\ug$ has burst-3 weight equal to 3 (in other
words, the minimal number of NAA bursts of length up to 3 that cover the
non-zero elements of $\ug$ is 3), so further tests are necessary on
$g(x)$ to determine the burst-correcting capability of the code. On
the other hand, if we would be testing if $g(x)$ generates a $[12,4,\lan
4,\ell\ran]$ code, $1\leq\ell\leq 4$, we can see that the
burst-4 weight of $\ug$ is 2, therefore no more tests are
necessary and we can move on to test the next polynomial.

So, let us go back to test if the code generated by $g(x)$ is a
$[12,4,\lan 3,\ell\ran]$ code. This one is a code of low dimension, so
basically we can check all possible codewords. A generator matrix for
the code is given by $\ug$ followed by 3 zeros and its 3 rotations to
the right, i.e.,

\begin{eqnarray*}
G&=&\left(
\begin{array}{cccccccccccc}
1&0&0&1&0&0&0&0&1&0&0&0\\
0&1&0&0&1&0&0&0&0&1&0&0\\
0&0&1&0&0&1&0&0&0&0&1&0\\
0&0&0&1&0&0&1&0&0&0&0&1\\
\end{array}\right)
\end{eqnarray*}

There are 15 possible non-zero codewords, but we can study the 8 that are
obtained by linear combinations containing the first row (the other 7
are rotations of those). These 8 codewords are:

$$
\begin{array}{cccccccccccc}
1&0&0&1&0&0&0&0&1&0&0&0\\
1&1&0&1&1&0&0&0&1&1&0&0\\
1&0&1&1&0&1&0&0&1&0&1&0\\
1&0&0&0&0&0&1&0&1&0&0&1\\
1&1&1&1&1&1&0&0&1&1&1&0\\
1&1&0&0&1&0&1&1&0&1&0&1\\
1&0&1&0&0&1&1&0&1&0&1&1\\
1&1&1&0&1&1&1&0&1&1&1&1\\
\end{array}
$$

All codewords need at least 3 bursts of length up to 3 to cover the
non-zero entries, except for

$$
(1\;0\;0\;0\;0\;0\;1\;0\;1\;0\;0\;1)\\
$$

This codeword can be covered by a NAA burst of length 3 together with
an AA
burst of length 2, therefore the code cannot be a $[12,4,\lan
3,2\ran]$ code. However, it is a $[12,4,\lan 3,1\ran]$ code. This
method is not practical when $k$ is large, since we would be
measuring the burst-distance of possibly half the codewords in the
code. For that reason, we describe next a method that is more
efficient for large $k$ and is based on the parity-check matrix of
the code as
opposed to the generator matrix.

We will again illustrate the algorithm with a particular example: we
will take $n\eq 14$, $b\eq 3$ and $g(x)\eq 1+x^3+x^4+x^5+x^6$. Since $g(x)$ has
degree 6, this is a $[14,8]$ code that we denote by $\C$. We will
explore whether  $\C$ is a (3,2) or a (3,3) single burst-correcting
code using the search algorithm to be described next(by the Reiger
bound, $\C$ cannot be a $(4,l)$ single burst-correcting code).

The first step is to obtain a generator matrix for $\C$ using $g(x)$.
This is very easily done, see for instance~\cite{b}. Explicitly,
since the code has dimension 8, the generator matrix $G$ is obtained
by shifting $g(x)$ in binary 8 times, i.e.,

\begin{eqnarray*}
G&=&\left(
\begin{array}{cccccccccccccc}
1&0&0&1&1&1&1&0&0&0&0&0&0&0\\
0&1&0&0&1&1&1&1&0&0&0&0&0&0\\
0&0&1&0&0&1&1&1&1&0&0&0&0&0\\
0&0&0&1&0&0&1&1&1&1&0&0&0&0\\
0&0&0&0&1&0&0&1&1&1&1&0&0&0\\
0&0&0&0&0&1&0&0&1&1&1&1&0&0\\
0&0&0&0&0&0&1&0&0&1&1&1&1&0\\
0&0&0&0&0&0&0&1&0&0&1&1&1&1\\
\end{array}\right)
\end{eqnarray*}

Now we want to obtain a parity-check matrix $H$ from $G$. In order to
do that, we need to put $G$ in systematic form, i.e., the first 8
columns of 8 need to be the identity. This is easily done from $G$ by
Gaussian elimination. Once the Gaussian elimination process is
finished, matrix $G$ is transformed into the systematic form

\begin{eqnarray*}
G_{\rm sys}&=&\left(
\begin{array}{cccccccccccccc}
1&0&0&0&0&0&0&0&1&1&0&1&0&0\\
0&1&0&0&0&0&0&0&0&1&1&0&1&0\\
0&0&1&0&0&0&0&0&0&0&1&1&0&1\\
0&0&0&1&0&0&0&0&1&0&0&0&0&1\\
0&0&0&0&1&0&0&0&1&1&0&1&1&1\\
0&0&0&0&0&1&0&0&1&1&1&1&0&0\\
0&0&0&0&0&0&1&0&0&1&1&1&1&0\\
0&0&0&0&0&0&0&1&0&0&1&1&1&1\\
\end{array}\right)
\end{eqnarray*}

When a systematic generator matrix has the form $(I_k\,|\,V)$, with
$I_k$ the $k\times k$ identity matrix and $V$ a $k\times (n-k)$
matrix, then a systematic parity-check matrix is given by
$H\eq (I_{n-k}\,|\,V^T)$, $V^T$ the transpose of $V$~\cite{b}.
Applying this formula to $G_{\rm sys}$ above, we obtain the
systematic parity-check matrix

\begin{eqnarray}
\label{eqH}
H&=&\left(
\begin{array}{cccccccccccccc}
1& 0& 0& 1& 1& 1& 0& 0& 1& 0& 0& 0& 0& 0\\
1& 1& 0& 0& 1& 1& 1& 0& 0& 1& 0& 0& 0& 0\\
0& 1& 1& 0& 0& 1& 1& 1& 0& 0& 1& 0& 0& 0\\
1& 0& 1& 0& 1& 1& 1& 1& 0& 0& 0& 1& 0& 0\\
0& 1& 0& 0& 1& 0& 1& 1& 0& 0& 0& 0& 1& 0\\
0& 0& 1& 1& 1& 0& 0& 1& 0& 0& 0& 0& 0& 1\\
\end{array}\right)
\end{eqnarray}

Assume that we want to check if the following double burst is a
codeword in the code defined by parity-check matrix $H$ above:

\begin{eqnarray*}
\uv_0&=&\left(
\begin{array}{cccccccccccccc}
1& 0& 1& 0& 0& 0& 0& 1& 1& 1& 0& 0& 0& 0\\
\end{array}\right)
\end{eqnarray*}

If it is a codeword, then the code is not a $[14,8,\lan
3,1\ran]$.
Since the code is shortened
cyclic, if $\uv_0$ is a codeword, then rotating it to the right is
also a codeword, due to the cyclic
property:

\begin{eqnarray*}
\uv_1&=&\left(
\begin{array}{cccccccccccccc}
0& 1& 0& 1& 0& 0& 0& 0& 1& 1& 1& 0& 0& 0\\
\end{array}\right)
\end{eqnarray*}

We can write $\uv_1\eq \uw_0\xor \uw_1$, where

\begin{eqnarray*}
\uw_0&=&\left(
\begin{array}{cccccccccccccc}
0& 1& 0& 1& 0& 0& 0& 0& 0& 0& 0& 0& 0& 0\\
\end{array}\right)\\
\uw_1&=&\left(
\begin{array}{cccccccccccccc}
0& 0& 0& 0& 0& 0& 0& 0& 1& 1& 1& 0& 0& 0\\
\end{array}\right)\\
\end{eqnarray*}

If $\uv_1$ is a codeword, both bursts of length 3 $\uw_0$ and
$\uw_1$, have the same syndrome.

In particular, the syndrome of $\uw_1$ is

\begin{eqnarray*}
\us_1&=&\uw_1 H^T\eq
\left(
\begin{array}{cccccc}
1& 1& 1& 0& 0& 0\\
\end{array}\right)\\
\end{eqnarray*}

In other words, the syndrome of a burst of length at most $b$
occurring in the last $n-k$
coordinates is also a (NAA) burst of length at most $b$ with respect to a
systematic parity-check matrix $H$. In the case of the parity-check
matrix given by~(\ref{eqH}), the list of syndromes of bursts of
length up to 3 occurring in the last 6 coordinates (we include the
all-zero syndrome) together with
their decimal representation is

\begin{center}
\begin{tabular}{|c|c|}
\hline
Syndrome&Decimal\\
\hline
$0\; 0\; 0\; 0\; 0\; 0$&0\\
$0\; 0\; 0\; 0\; 0\; 1$&1\\
$0\; 0\; 0\; 0\; 1\; 0$&2\\
$0\; 0\; 0\; 0\; 1\; 1$&3\\
\hline
$0\; 0\; 0\; 1\; 0\; 0$&4\\
$0\; 0\; 0\; 1\; 0\; 1$&5\\
$0\; 0\; 0\; 1\; 1\; 0$&6\\
$0\; 0\; 0\; 1\; 1\; 1$&7\\
\hline
$0\; 0\; 1\; 0\; 0\; 0$&8\\
$0\; 0\; 1\; 0\; 1\; 0$&10\\
$0\; 0\; 1\; 1\; 0\; 0$&12\\
$0\; 0\; 1\; 1\; 1\; 0$&14\\
\hline
$0\; 1\; 0\; 0\; 0\; 0$&16\\
$0\; 1\; 0\; 1\; 0\; 0$&20\\
$0\; 1\; 1\; 0\; 0\; 0$&24\\
$0\; 1\; 1\; 1\; 0\; 0$&28\\
\hline
$1\; 0\; 0\; 0\; 0\; 0$&32\\
$1\; 0\; 1\; 0\; 0\; 0$&40\\
$1\; 1\; 0\; 0\; 0\; 0$&48\\
$1\; 1\; 1\; 0\; 0\; 0$&56\\
\hline
\end{tabular}
\end{center}

In general, there are $2^{b-1}(n-k-(b-2))$ NAA bursts of length up
to $b$ among the vectors of length $n-k$ corresponding to the
syndromes, where $2b\leq n-k$ by the Reiger bound. In decimal, the
corresponding numbers are $0\leq i\leq 2^{b-1}-1$, and
$2^jt$, where $2^{b-1}\leq t\leq 2^{b}-1$ and $0\leq j\leq
n-k-b$.

Remember that we want to check if the code is $[n,k,\lan b,\ell\ran]$.
If $\ell>1$, we need to
compute the syndromes of all the all-around bursts of length up to
$\ell$ and store them (one way of storing them is by writing them in decimal).
If one of the syndromes of the all-around bursts is a NAA burst of
length at most $b$, then the code is not $[n,k,\lan b,\ell\ran]$. Let us
illustrate the concepts by going back to our
example with the $[14,8]$
code as given by parity-check matrix $H$ in~(\ref{eqH}) to show how this is done.

Consider $\ell\eq 2$. This case is simple since there is
only one all-around burst of length exactly 2, which is

\begin{eqnarray*}
\uw_4&=&\left(
\begin{array}{cccccccccccccc}
1& 0& 0& 0& 0& 0& 0& 0& 0& 0& 0& 0& 0& 1\\
\end{array}\right)\\
\end{eqnarray*}

The syndrome of this burst is

\begin{eqnarray*}
\us_4&=&\uw_4 H^T\eq
\left(
\begin{array}{cccccc}
1& 1& 0& 1& 0& 1\\
\end{array}\right)\\
\end{eqnarray*}
(it is the XOR of the first and last columns of $H$), which in
decimal is 53.

If we take $\ell\eq 3$, we have to find the syndromes of all the
all-around bursts of length up to 3 (in decimal). We have already
found the syndrome corresponding to the only all-around burst of
length 2, let us find now the syndromes corresponding to all-around
bursts of length exactly 3. The all-around bursts of length exactly 3
are:

\begin{eqnarray*}
\uw_5&=&\left(
\begin{array}{cccccccccccccc}
1& 0& 0& 0& 0& 0& 0& 0& 0& 0& 0& 0& 1& 0\\
\end{array}\right)\\
\uw_6&=&\left(
\begin{array}{cccccccccccccc}
1& 0& 0& 0& 0& 0& 0& 0& 0& 0& 0& 0& 1& 1\\
\end{array}\right)\\
\uw_7&=&\left(
\begin{array}{cccccccccccccc}
0& 1& 0& 0& 0& 0& 0& 0& 0& 0& 0& 0& 0& 1\\
\end{array}\right)\\
\uw_8&=&\left(
\begin{array}{cccccccccccccc}
1& 1& 0& 0& 0& 0& 0& 0& 0& 0& 0& 0& 0& 1\\
\end{array}\right)\\
\end{eqnarray*}

The corresponding syndromes are

\begin{eqnarray*}
\us_5&=&\uw_5 H^T\eq
\left(
\begin{array}{cccccc}
1& 1& 0& 1& 1& 0\\
\end{array}\right)\\
\us_6&=&\uw_6 H^T\eq
\left(
\begin{array}{cccccc}
1& 1& 0& 1& 1& 1\\
\end{array}\right)\\
\us_7&=&\uw_7 H^T\eq
\left(
\begin{array}{cccccc}
0& 1& 1& 0& 1& 1\\
\end{array}\right)\\
\us_8&=&\uw_8 H^T\eq
\left(
\begin{array}{cccccc}
1& 0& 1& 1& 1& 1\\
\end{array}\right)\\
\end{eqnarray*}
In decimal, these four numbers correspond to 54, 55, 27 and 47 respectively.
For general $\ell$, the number of all-around bursts of length exactly
$\ell$ is $(\ell-1)2^{\ell-2}$. Therefore, if $l\eq 2$, the number is 1, while
if it is $\ell\eq 3$ it is 4; that was the case in our examples. If
$\ell\eq 4$ then we would have 12 all-around bursts of length exactly 4,
and so on.

Let us go back to our $[14,8]$ code generated by $g(x)$ and let's see
if it is a (3,3) single burst-correcting code. The set of syndromes
(in decimal) corresponding to NAA bursts of length up to 3 occurring
in the last 6 (in general, $n-k$) coordinates together with those
corresponding to
all-around bursts of
length up to 3, we have seen, is

\begin{eqnarray}
\label{eqS}
S&=&\{1,2,3,4,5,6,7,8,10,12,14,16,20,24,28,32,40,48,56,27,47,53,54,55\}
\end{eqnarray}

Now we have to examine the syndromes (in decimal) of all possible NAA
bursts of length up to 3 starting in coordinates 0 to 7 (in general,
0 to $k-1$) and check if they are
in set $S$. If there is at least one syndrome in set $S$, the code is
not (3,3) single burst-correcting, otherwise it is.

Next we describe an efficient method for computing the syndromes of
such NAA bursts of
length up to 3 by using Gray codes~\cite{s}. The most commonly used
construction of Gray codes is the reflective construction, and is the
one we are going to use here. Let us recall the construction of the
usual reflective Gray code. We proceed by
induction, giving explicitly the $2^m\times m$ matrix $\G(m)$ of the
Gray code of vectors of length $m$. Given a matrix $V$, denote by
$\tilde{V}$ the matrix $V$ with its rows in reverse order. For
example, if

\begin{eqnarray*}
V&=&\left(
\begin{array}{ccc}
1&0&1\\
0&1&1
\end{array}\right),
\end{eqnarray*}
then

\begin{eqnarray*}
\tilde{V}&=&\left(
\begin{array}{ccc}
0&1&1\\
1&0&1
\end{array}\right).
\end{eqnarray*}

So, for $m=1$, define

\begin{eqnarray*}
\G(1)&=&\left(
\begin{array}{c}
0\\
1
\end{array}\right),
\end{eqnarray*}
and for $m\geq 2$, let

\begin{eqnarray*}
\G(m)&=&\left(
\begin{array}{c|c}
\begin{array}{c}
0\\\vdots \\ 0
\end{array}
&
\G(m-1)
\\
\hline
\begin{array}{c}
1\\\vdots \\ 1
\end{array}&
\tilde{\G}(m-1)\end{array}
\right).
\end{eqnarray*}

For example,

\begin{eqnarray*}
\G(2)&=&\left(
\begin{array}{cc}
0&0\\
0&1\\
1&1\\
1&0\\
\end{array}\right),
\end{eqnarray*}

We have to compute the syndromes of 32 bursts of length at most 3,
starting in locations from 0 to 7 (in general, from 0 to $k-1$). We
will compute such 32 syndromes (in general, $k2^{b-1}$). At each
step, we XOR one column of $H$ with the
previously found syndrome, minimizing the number of operations. We
will use the $\G(2)$ Gray code found above to this end. One way to
write such 32 bursts is as follows:

$$
\begin{array}{llllllllllcccc}
0\phantom{1}&1\phantom{1}&2\phantom{1}&3\phantom{1}&4\phantom{1}
&5\phantom{1}&6\phantom{1}&7\phantom{1}&8\phantom{1}&9\phantom{1}
&10&11&12&13\\
\hline
1&0&0&0&0&0&0&0&0&0&0&0&0&0\\
1&0&1&0&0&0&0&0&0&0&0&0&0&0\\
1&1&1&0&0&0&0&0&0&0&0&0&0&0\\
1&1&0&0&0&0&0&0&0&0&0&0&0&0\\
\hline
0&1&0&0&0&0&0&0&0&0&0&0&0&0\\
0&1&0&1&0&0&0&0&0&0&0&0&0&0\\
0&1&1&1&0&0&0&0&0&0&0&0&0&0\\
0&1&1&0&0&0&0&0&0&0&0&0&0&0\\
\hline
0&0&1&0&0&0&0&0&0&0&0&0&0&0\\
0&0&1&0&1&0&0&0&0&0&0&0&0&0\\
0&0&1&1&1&0&0&0&0&0&0&0&0&0\\
0&0&1&1&0&0&0&0&0&0&0&0&0&0\\
\hline
0&0&0&1&0&0&0&0&0&0&0&0&0&0\\
0&0&0&1&0&1&0&0&0&0&0&0&0&0\\
0&0&0&1&1&1&0&0&0&0&0&0&0&0\\
0&0&0&1&1&0&0&0&0&0&0&0&0&0\\
\hline
0&0&0&0&1&0&0&0&0&0&0&0&0&0\\
0&0&0&0&1&0&1&0&0&0&0&0&0&0\\
0&0&0&0&1&1&1&0&0&0&0&0&0&0\\
0&0&0&0&1&1&0&0&0&0&0&0&0&0\\
\hline
0&0&0&0&0&1&0&0&0&0&0&0&0&0\\
0&0&0&0&0&1&0&1&0&0&0&0&0&0\\
0&0&0&0&0&1&1&1&0&0&0&0&0&0\\
0&0&0&0&0&1&1&0&0&0&0&0&0&0\\
\hline
0&0&0&0&0&0&1&0&0&0&0&0&0&0\\
0&0&0&0&0&0&1&0&1&0&0&0&0&0\\
0&0&0&0&0&0&1&1&1&0&0&0&0&0\\
0&0&0&0&0&0&1&1&0&0&0&0&0&0\\
\hline
0&0&0&0&0&0&0&1&0&0&0&0&0&0\\
0&0&0&0&0&0&0&1&0&1&0&0&0&0\\
0&0&0&0&0&0&0&1&1&1&0&0&0&0\\
0&0&0&0&0&0&0&1&1&0&0&0&0&0\\
\end{array}
$$

Take the first 4 bursts above. Coordinate 0 is always 1. The next two
coordinates, i.e., coordinates 1 and 2, constitute the Gray code
$\G(2)$ (in general, we use the Gray code $\G(b-1)$). When
we reach the 5th burst, coordinate 0 becomes 0. And we repeat the
process until covering all possible 32 bursts. The idea is to modify
only one coordinate at a time when we move from one burst to the
next. Let us illustrate how to compute the syndromes by using the
parity-check matrix given by~(\ref{eqH}). Let us rewrite the columns
of $H$ with their corresponding location numbers:

\begin{eqnarray*}
H&=&\left(
\begin{array}{llllllllllcccc}
0\phantom{1}&1\phantom{1}&2\phantom{1}&3\phantom{1}&4\phantom{1}
&5\phantom{1}&6\phantom{1}&7\phantom{1}&8\phantom{1}&9\phantom{1}
&10&11&12&13\\
\hline
1& 0& 0& 1& 1& 1& 0& 0& 1& 0& 0& 0& 0& 0\\
1& 1& 0& 0& 1& 1& 1& 0& 0& 1& 0& 0& 0& 0\\
0& 1& 1& 0& 0& 1& 1& 1& 0& 0& 1& 0& 0& 0\\
1& 0& 1& 0& 1& 1& 1& 1& 0& 0& 0& 1& 0& 0\\
0& 1& 0& 0& 1& 0& 1& 1& 0& 0& 0& 0& 1& 0\\
0& 0& 1& 1& 1& 0& 0& 1& 0& 0& 0& 0& 0& 1\\
\end{array}\right)
\end{eqnarray*}

The syndrome of the first burst is given by column 0 of $H$,
which as a vector is \\$\us\eq (1\;1\;0\;1\;0\;0)$, and in decimal
corresponds number 52. This number is not in $S$ as given by~(\ref{eqS}), so we
continue with our search. For the syndrome of the second burst we XOR
$\us$ with column 2 in $H$, giving
$\us\eq (1\;1\;1\;0\;0\;1)$, which in binary is 57. Again, this number is
not in $S$. Next we XOR $\us$ with column 1 in $H$, giving
$\us\eq (1\;0\;0\;0\;1\;1)$, which in binary is 35, not in the list.
Next we XOR $\us$ with column 2 in $H$, giving
$\us\eq (1\;0\;1\;1\;1\;0)$, which in binary is 46, not in the list.
Next we XOR $\us$ with column 0 in $H$, giving
$\us\eq (0\;1\;1\;0\;1\;0)$, which in binary is 26, not in the list.
We continue this way the process. A way to describe the computation
of the 32 syndromes is as follows: take as initial syndrome $\us$
column 0 of $H$ and check whether it is in list $S$ or not. If it is, the
code is not $[14,8,\lan 3,3\ran]$. If it is not, continue. At each
step, continue computing the syndrome by XORing the previously
computed syndrome with columns 2, 1, 2, 0, 3, 2, 3, 1, 4, 3, 4, 2, 5,
4, 5, 3, 6, 5, 6, 4, 7, 6, 7, 5, 8, 7, 8, 6, 9, 8, 9. At each step we
verify whether
$\us$ in decimal is in $S$ or not. Let us continue like this. Then, we
verify that the successive syndromes, illustrated in the following
table, are (convention, assume that $\uh_i$, $0\leq i\leq 13$,
represent the columns of $H$ as horizontal vectors and initially
$\us\eq 0\;0\;0\;0\;0\;0$):

$$
\begin{array}{|c|c|c|c|}
\hline
{\rm Burst}&j&\us\eq \us \xor \uh_j&{\rm Decimal}\\
\hline
1\;0\;0\;0\;0\;0\;0\;0\;0\;0\;0\;0\;0\;0&0 &1\;1\;0\;1\;0\;0&52\\
1\;0\;1\;0\;0\;0\;0\;0\;0\;0\;0\;0\;0\;0&2 &1\;1\;0\;0\;1\;0&57\\
1\;1\;1\;0\;0\;0\;0\;0\;0\;0\;0\;0\;0\;0&1 &1\;0\;0\;0\;1\;1&35\\
1\;1\;0\;0\;0\;0\;0\;0\;0\;0\;0\;0\;0\;0&2 &1\;0\;1\;1\;1\;0&46\\
0\;1\;0\;0\;0\;0\;0\;0\;0\;0\;0\;0\;0\;0&0 &0\;1\;1\;0\;1\;0&26\\
0\;1\;0\;1\;0\;0\;0\;0\;0\;0\;0\;0\;0\;0&3 &1\;1\;1\;0\;1\;1&59\\
0\;1\;1\;1\;0\;0\;0\;0\;0\;0\;0\;0\;0\;0&2 &1\;1\;1\;1\;0\;1&54\\
0\;1\;1\;0\;0\;0\;0\;0\;0\;0\;0\;0\;0\;0&3 &0\;1\;0\;1\;1\;1&23\\
0\;0\;1\;0\;0\;0\;0\;0\;0\;0\;0\;0\;0\;0&1 &0\;0\;1\;1\;0\;1&13\\
0\;0\;1\;0\;1\;0\;0\;0\;0\;0\;0\;0\;0\;0&4 &1\;1\;1\;0\;1\;0  &58\\
0\;0\;1\;1\;1\;0\;0\;0\;0\;0\;0\;0\;0\;0&3 &0\;1\;1\;0\;1\;1  &27\\
0\;0\;1\;1\;0\;0\;0\;0\;0\;0\;0\;0\;0\;0&4 &1\;0\;1\;1\;0\;0  &44\\
0\;0\;0\;1\;0\;0\;0\;0\;0\;0\;0\;0\;0\;0&2 &1\;0\;0\;0\;0\;1  &33\\
0\;0\;0\;1\;0\;1\;0\;0\;0\;0\;0\;0\;0\;0&5 &0\;1\;1\;1\;0\;1  &29\\
0\;0\;0\;1\;1\;1\;0\;0\;0\;0\;0\;0\;0\;0&4 &1\;0\;1\;0\;1\;0  &42\\
0\;0\;0\;1\;1\;0\;0\;0\;0\;0\;0\;0\;0\;0&5 &0\;1\;0\;1\;1\;0  &22\\
0\;0\;0\;0\;1\;0\;0\;0\;0\;0\;0\;0\;0\;0&3 &1\;1\;0\;1\;1\;1  &55\\
\hline
\end{array}
$$

Notice that syndrome 55, corresponding to the last burst above, is in
set $S$. We have seen that this
syndrome also corresponds to the AA burst of length 3

\begin{eqnarray*}
\uw_6&=&\left(
\begin{array}{cccccccccccccc}
1& 0& 0& 0& 0& 0& 0& 0& 0& 0& 0& 0& 1& 1\\
\end{array}\right)\\
\end{eqnarray*}

So, $\C$
is not a $[14,8,\lan 3,3\ran]$ code.
Is $\C$ a $[14,8,\lan 3,2\ran]$ code? In this case the set
of syndromes is reduced. Explicitly, it is given by

\begin{eqnarray*}
S'&=&\{1,2,3,4,5,6,7,8,10,12,14,16,20,24,28,32,40,48,56,53\}\\
\end{eqnarray*}

Continuing with the search, we obtain

$$
\begin{array}{|c|r|c|c|}
\hline
{\rm Burst}&j&\us\eq\us \xor \uh_j&{\rm Decimal}\\
\hline
0\;0\;0\;0\;1\;0\;0\;0\;0\;0\;0\;0\;0\;0&3 &1\;1\;0\;1\;1\;1  &55\\
0\;0\;0\;0\;1\;0\;1\;0\;0\;0\;0\;0\;0\;0&6 &1\;0\;1\;0\;0\;1  &41\\
0\;0\;0\;0\;1\;1\;1\;0\;0\;0\;0\;0\;0\;0&5 &0\;1\;0\;1\;0\;1  &21\\
0\;0\;0\;0\;1\;1\;0\;0\;0\;0\;0\;0\;0\;0&6 &0\;0\;1\;0\;1\;1  &11\\
0\;0\;0\;0\;0\;1\;0\;0\;0\;0\;0\;0\;0\;0&4 &1\;1\;1\;1\;0\;0  &60\\
0\;0\;0\;0\;0\;1\;0\;1\;0\;0\;0\;0\;0\;0&7 &1\;1\;0\;0\;1\;1  &51\\
0\;0\;0\;0\;0\;1\;1\;1\;0\;0\;0\;0\;0\;0&6 &1\;0\;1\;1\;0\;1  &45\\
0\;0\;0\;0\;0\;1\;1\;0\;0\;0\;0\;0\;0\;0&7 &1\;0\;0\;0\;1\;0  &34\\
0\;0\;0\;0\;0\;0\;1\;0\;0\;0\;0\;0\;0\;0&5 &0\;1\;1\;1\;1\;0  &30\\
0\;0\;0\;0\;0\;0\;1\;0\;1\;0\;0\;0\;0\;0&8 &1\;1\;1\;1\;1\;0  &62\\
0\;0\;0\;0\;0\;0\;1\;1\;1\;0\;0\;0\;0\;0&7 &1\;1\;0\;0\;0\;1  &49\\
0\;0\;0\;0\;0\;0\;1\;1\;0\;0\;0\;0\;0\;0&8 &0\;1\;0\;0\;0\;1  &17\\
0\;0\;0\;0\;0\;0\;0\;1\;0\;0\;0\;0\;0\;0&6 &0\;0\;1\;1\;1\;1  &15\\
0\;0\;0\;0\;0\;0\;0\;1\;0\;1\;0\;0\;0\;0&9 &0\;1\;1\;1\;1\;1  &31\\
0\;0\;0\;0\;0\;0\;0\;1\;1\;1\;0\;0\;0\;0&8 &1\;1\;1\;1\;1\;1  &63\\
0\;0\;0\;0\;0\;0\;0\;1\;1\;0\;0\;0\;0\;0&9 &1\;0\;1\;1\;1\;1  &47\\
\hline
\end{array}
$$

We can see that none of the syndromes is in $S'$, so $\C$ is a (3,2)
single burst-correcting code.

The general case is analogous to the one illustrated in this example.
We are ready to state the search algorithm which is the main result
of this paper.

\begin{alg}
{\rm
Given $n$, $k$, $b\leq (n-k)/2$ and $1\leq \ell\leq b$, the algorithm finds out
if there is a cyclic
or shortened cyclic $[n,k,\lan b,\ell\ran]$ code $\C$ with generator
polynomial $g(x)\eq g_0+g_1x+\cdots +g_{n-k}x^{n-k}$, $g_0\eq g_{n-k}\eq 1$.
The candidate polynomials are examined in alphabetical order. Let $\ug\eq
(g_0,g_1,\ldots,g_{n-k})$ and $\stackrel{\la}{\underline{g}}\eq
(g_{n-k},g_{n-k-1},\ldots,g_0)$. Then:

\begin{enumerate}

\item If $\ug\eq (1,1,\ldots,1)$ declare that there is no $[n,k,\lan
b,\ell\ran]$ code $\C$ and exit.

\item If $g_b\eq g_{b+1}\eq \cdots\eq g_{n-k-b}\eq 0$, then consider
the next $\ug$ in lexicographic order and go to step 1.

\item If $\ug\,>\,\stackrel{\la}{\underline{g}}$, then move to
the next $\ug$ and go to step 1, where we consider the relationship
'$>$' in lexicographic order.

\item Consider the $k\times n$ generator matrix

\begin{eqnarray*}
G&=&\left(
\begin{array}{ccccccccc}
g_0&g_1&g_2&\ldots &g_{n-k}&0&0&\ldots &0\\
  0&g_0&g_1&\ldots &g_{n-k-1}&g_{n-k}&0&\ldots &0\\
\vdots&\vdots&\vdots&\ddots &\vdots&\vdots&\vdots&\ddots &\vdots\\
  0&0&0&\ldots &0&g_0&g_1&\ldots &g_{n-k}\\
\end{array}
\right)
\end{eqnarray*}

\item By Gaussian elimination on $G$, obtain the systematic generator
matrix $G_{\rm sys}\eq (I_k\,|\,V)$, where $I_k$ is the $k\times k$
identity matrix and $V$ is a $k\times (n-k)$ matrix.

\item From $G_{\rm sys}$, obtain the systematic parity-check matrix
$H\eq (V^T\,|\,I_{n-k})$. Denote by $\uh_0,\uh_1,\ldots,\uh_{k-1}$ the
first $k$ columns of $H$.

\item Consider the
$2^{b-1}(n-k-(b-2))$
numbers $0\leq i\leq 2^{b-1}-1$, and
$2^jt$, where \\$2^{b-1}\leq t\leq 2^{b}-1$ and $0\leq j\leq
n-k-b$, corresponding to the syndromes of the\\ $2^{b-1}(n-k-(b-2))$
NAA bursts of length up to $b$ whose first $k$ coordinates are 0
(including the all-zero vector).
Consider also the $(\ell -2)2^{\ell -1}+1$ numbers corresponding to
the $(\ell -2)2^{\ell -1}+1$ syndromes of AA bursts of length up to
$\ell$. If one of these numbers gets repeated, then consider the next
$g(x)$ in lexicographic order and go to step 1. Otherwise, call $S$
the set consisting of these
$2^{b-1}(n-k-(b-2))+(\ell -2)2^{\ell -1}+1$ numbers.

\item Consider a reflective Gray code $\G(b-1)$.
Let $j\la 0$, $\us$ the all-zero vector of length $n-k$ and $s\eq 0$.

\item Let $j\eq q\,2^{b-1}+t$, with $0\leq t< 2^{b-1}$. If
$t\eq 0$, then let $\us\la \us\xor \uc_q$. If $t\neq 0$, let $\us\la
\us\xor \uc_{q+d}$, where $d$ is the coordinate changing between rows
$t-1$ and $t$ of the Gray code \\ $\G(b-1)$. Let $s$ be the decimal
representation of $\us$. If $s\in S$, then consider the next $g(x)$ in
lexicographic order and go back to step 1. Otherwise make $j\la j+1$.

\item If $j\eq k-1$, then declare that code $\C$ generated by $g(x)$
is an $[n,k,\lan b,\ell\ran]$ code and exit. Otherwise go back to
step 9.

\end{enumerate}
\qed

}
\end{alg}

In step 2 of the algorithm above we are checking whether $g(x)$ as a
binary vector has burst weight larger than 2. In step 3 we avoid
checking polynomials that we have already checked in reverse order.

Next we present tables with the best parameters for different
values of burst and guard space lengths.

\section{Tables obtained for different pairs $(b,g)$}
\label{tables}

In the following pages tables with optimum single
burst-correcting-codes for $5{\ }{\leq}{\ }b{\ }{\leq}{\ }10$ and for
$20{\ }{\leq}{\ }g{\ }{\leq}{\ }100$ are presented. The tables were
obtained using Algorithm~2.1. The
generator polynomials are given in hexadecimal. In each row, the
code giving the best rate for that particular guard space is framed.

\begin{table}
\begin{center}
\tiny{
\begin{tabular}{|c|c|c|c|c|c|c|c|c|}
\hline
$g$ &$5_1$ &$5_2$ &$5_3$ &$5_4$ &$5_5$&max $k/n$& Polynomial& Cyclic?\\
\hline
20& [21,11]& [22,12]& $\fbox{[23,13]}$& [24,13]& [25,14]& [23,13]& 4ED& No\\
21& [22,12]& $\fbox{[23,13]}$& [24,13]& [25,14]& [26,14]& [23,13]& 4ED& No\\
22& [23,13]& [24,14]& $\fbox{[25,15]}$& [26,15]& [27,15]& [25,15]& 523& No\\
23& [24,14]& $\fbox{[25,15]}$& [26,15]& [27,16]& [28,16]& [25,15]& 4ED& No\\
24& [25,15]& [26,16]& $\fbox{[27,17]}$& [28,17]& [29,17]& [27,17]& 523& No\\
25& [26,16]& $\fbox{[27,17]}$& [28,17]& [29,18]& [30,18]& [27,17]& 4ED& No\\
26& [27,17]& [28,17]& [29,18]& [30,19]& $\fbox{[31,20]}$& [31,20]& 867& Yes\\
27& [28,17]& [29,18]& [30,19]& $\fbox{[31,20]}$& [32,20]& [31,20]& 867& Yes\\
28& [29,18]& [30,19]& [31,20]& $\fbox{[32,21]}$& [33,21]& [32,21]& 947& No\\
29& [30,19]& [31,20]& [32,21]& $\fbox{[33,22]}$& [34,22]& [33,22]& 947& No\\
30& [31,20]& [32,21]& [33,22]& $\fbox{[34,23]}$& [35,23]& [34,23]& 837& No\\
31& [32,21]& [33,22]& [34,23]& $\fbox{[35,24]}$& [36,24]& [35,24]& ABD& No\\
32& [33,22]& [34,23]& [35,24]& $\fbox{[36,25]}$& [37,25]& [36,25]& 83D& No\\
33& [34,23]& [35,24]& [36,25]& $\fbox{[37,26]}$& [38,26]& [37,26]& 9CD& No\\
34& [35,24]& [36,25]& $\fbox{[37,26]}$& [38,26]& [39,27]& [37,26]& 83D& No\\
35& [36,25]& [37,26]& $\fbox{[38,27]}$& [39,27]& [40,28]& [38,27]& 8B7& No\\
36& [37,26]& [38,27]& $\fbox{[39,28]}$& [40,28]& [41,29]& [39,28]& A6D& No\\
37& [38,27]& [39,28]& $\fbox{[40,29]}$& [41,29]& [42,30]& [40,29]& 8D3& No\\
38& [39,28]& [40,29]& $\fbox{[41,30]}$& [42,30]& [43,31]& [41,30]& DA7& No\\
39& [40,29]& [41,30]& $\fbox{[42,31]}$& [43,31]& [44,31]& [42,31]& BEF& No\\
40& [41,30]& [42,31]& $\fbox{[43,32]}$& [44,32]& [45,33]& [43,32]& 8BF& No\\
41& [42,31]& [43,32]& $\fbox{[44,33]}$& [45,33]& [46,33]& [44,33]& 829& No\\
42& [43,32]& [44,33]& [45,34]& $\fbox{[46,35]}$& [47,35]& [46,35]& 8BF& No\\
43& [44,33]& [45,34]& [46,35]& $\fbox{[47,36]}$& [48,35]& [47,36]& 8BF& No\\
44& [45,34]& [46,35]& $\fbox{[47,36]}$& [48,36]& [49,36]& [47,36]& 8BF& No\\
45& [46,35]& $\fbox{[47,36]}$& [48,36]& [49,37]& [50,37]& [47,36]& 8BF& No\\
46& $\fbox{[47,36]}$& [48,36]& [49,37]& [50,38]& [51,39]& [47,36]& 829& No\\
47& $\fbox{[48,37]}$& [49,37]& [50,38]& [51,39]& [52,39]& [48,37]& 829& No\\
48& [49,37]& [50,38]& [51,39]& $\fbox{[52,40]}$& [53,40]& [52,40]& 1021& No\\
49& [50,38]& [51,39]& [52,40]& $\fbox{[53,41]}$& [54,41]& [53,41]& 1021& No\\
50& [51,39]& [52,40]& [53,41]& $\fbox{[54,42]}$& [55,42]& [54,42]& 116D& No\\
51& [52,40]& [53,41]& [54,42]& $\fbox{[55,43]}$& [56,43]& [55,43]& 1245& No\\
52& [53,41]& [54,42]& [55,43]& $\fbox{[56,44]}$& [57,44]& [56,44]& 1147& No\\
53& [54,42]& [55,43]& [56,44]& $\fbox{[57,45]}$& [58,45]& [57,45]& 1059& No\\
54& [55,43]& [56,44]& [57,45]& [58,46]& $\fbox{[59,47]}$& [59,47]& 1AD7& No\\
55& [56,44]& [57,45]& [58,46]& $\fbox{[59,47]}$& [60,47]& [59,47]& 11BD& No\\
56& [57,45]& [58,46]& [59,47]& $\fbox{[60,48]}$& [61,48]& [60,48]& 10CD& No\\
57& [58,46]& [59,47]& [60,48]& $\fbox{[61,49]}$& [62,49]& [61,49]& 106F& No\\
58& [59,47]& [60,48]& [61,49]& [62,50]& $\fbox{[63,51]}$& [63,51]& 105F& Yes\\
59& [60,48]& [61,49]& [62,50]& [63,51]& $\fbox{[64,52]}$& [64,52]& 1237& No\\
60& [61,49]& [62,50]& [63,51]& $\fbox{[64,52]}$& [65,52]& [64,52]& 1237& No\\
\hline
\end{tabular}
}
\end{center}
\begin{center}
\caption{Optimal (shortened) cyclic codes correcting bursts of length up
to 5, for a guard\newline space from $g = 20$ to $g = 60$}
\end{center}
\end{table}

\begin{table}
\begin{center}
\tiny{
\begin{tabular}{|c|c|c|c|c|c|c|c|c|}
\hline
$g$ &$5_1$ &$5_2$ &$5_3$ &$5_4$ &$5_5$&max $k/n$& Polynomial& Cyclic?\\
\hline
61& [62,50]& [63,51]& [64,52]& $\fbox{[65,53]}$& [66,53]& [65,53]& 16CF& No\\
62& [63,51]& [64,52]& [65,53]& $\fbox{[66,54]}$& [67,54]& [66,54]& 1235& No\\
63& [64,52]& [65,53]& [66,54]& $\fbox{[67,55]}$& [68,55]& [67,55]& 1235& No\\
64& [65,53]& [66,54]& [67,55]& $\fbox{[68,56]}$& [69,56]& [68,56]& 142D& No\\
65& [66,54]& [67,55]& [68,56]& $\fbox{[69,57]}$& [70,57]& [69,57]& 142D& No\\
66& [67,55]& [68,56]& [69,57]& $\fbox{[70,58]}$& [71,58]& [70,58]& 1075& No\\
67& [68,56]& [69,57]& $\fbox{[70,58]}$& [71,58]& [72,59]& [70,58]& 106F& No\\
68& [69,57]& [70,58]& [71,59]& $\fbox{[72,60]}$& [73,60]& [72,60]& 1075& No\\
69& [70,58]& [71,59]& [72,60]& $\fbox{[73,61]}$& [74,61]& [73,61]& 1BBF& No\\
70& [71,59]& [72,60]& [73,61]& $\fbox{[74,62]}$& [75,62]& [74,62]& 1BBF& No\\
71& [72,60]& [73,61]& $\fbox{[74,62]}$& [75,62]& [76,63]& [74,62]& 102B& No\\
72& [73,61]& [74,62]& [75,63]& $\fbox{[76,64]}$& [77,64]& [76,64]& 1747& No\\
73& [74,62]& [75,63]& [76,64]& $\fbox{[77,65]}$& [78,65]& [77,65]& 18D3& No\\
74& [75,63]& [76,64]& [77,65]& $\fbox{[78,66]}$& [79,66]& [78,66]& 106F& No\\
75& [76,64]& [77,65]& [78,66]& $\fbox{[79,67]}$& [80,67]& [79,67]& 1423& No\\
76& [77,65]& [78,66]& [79,67]& $\fbox{[80,68]}$& [81,68]& [80,68]& 1423& No\\
77& [78,66]& [79,67]& $\fbox{[80,68]}$& [81,68]& [82,69]& [80,68]& 1423& No\\
78& [79,67]& [80,68]& [81,69]& $\fbox{[82,70]}$& [83,70]& [82,70]& 17B7& No\\
79& [80,68]& [81,69]& $\fbox{[82,70]}$& [83,70]& [84,71]& [82,70]& 102B& No\\
80& [81,69]& [82,70]& [83,71]& [84,72]& $\fbox{[85,73]}$& [85,73]& 1059& Yes\\
81& [82,70]& [83,71]& [84,72]& $\fbox{[85,73]}$& [86,73]& [85,73]& 1059& Yes\\
82& [83,71]& [84,72]& [85,73]& $\fbox{[86,74]}$& [87,74]& [86,74]& 1255& No\\
83& [84,72]& [85,73]& [86,74]& $\fbox{[87,75]}$& [88,74]& [87,75]& 1255& No\\
84& [85,73]& [86,74]& [87,75]& [88,76]& $\fbox{[89,77]}$& [89,77]& 1453& Yes\\
85& [86,74]& [87,75]& [88,76]& $\fbox{[89,77]}$& [90,77]& [89,77]& 1453& Yes\\
86& [87,75]& [88,76]& $\fbox{[89,77]}$& [90,77]& [91,78]& [89,77]& 1453& Yes\\
87& [88,76]& [89,77]& $\fbox{[90,78]}$& [91,78]& [92,79]& [90,78]& 1175& No\\
88& [89,77]& [90,78]& [91,79]& [92,80]& $\fbox{[93,81]}$& [93,81]& 1175& Yes\\
89& [90,78]& [91,79]& [92,80]& $\fbox{[93,81]}$& [94,81]& [93,81]& 1175& Yes\\
90& [91,79]& [92,80]& $\fbox{[93,81]}$& [94,81]& [95,82]& [93,81]& 1175& Yes\\
91& [92,80]& [93,81]& $\fbox{[94,82]}$& [95,82]& [96,83]& [94,82]& 1733& No\\
92& [93,81]& $\fbox{[94,82]}$& [95,82]& [96,83]& [97,83]& [94,82]& 106F& No\\
93& [94,82]& $\fbox{[95,83]}$& [96,83]& [97,84]& [98,85]& [95,83]& 13A3& No\\
94& [95,83]& [96,84]& $\fbox{[97,85]}$& [98,85]& [99,86]& [97,85]& 1733& No\\
95& [96,84]& $\fbox{[97,85]}$& [98,85]& [99,86]& [100,87]& [97,85]& 106F& No\\
96& [97,85]& $\fbox{[98,86]}$& [99,86]& [100,87]& [101,88]& [98,86]& 116D& No\\
97& [98,86]& [99,87]& $\fbox{[100,88]}$& [101,88]& [102,88]& [100,88]& 19CB& No\\
98& [99,87]& [100,88]& [101,89]& $\fbox{[102,90]}$& [103,90]& [102,90]& 1BBF& No\\
99& [100,88]& [101,89]& [102,90]& $\fbox{[103,91]}$& [104,91]& [103,91]& 1BBF& No\\
100& [101,89]& [102,90]& [103,91]& [104,92]& $\fbox{[105,93]}$& [105,93]& 116D& Yes\\
\hline
\end{tabular}
}
\end{center}
\begin{center}
\caption{Optimal (shortened) cyclic codes correcting bursts of length up
to 5, for a guard\newline space from $g = 61$ to $g = 100$}
\end{center}
\end{table}

\begin{table}
\begin{center}
\tiny{
\begin{tabular}{|c|c|c|c|c|c|c|c|c|}
\hline
$g$ &$6_1$ &$6_2$ &$6_3$ &$6_4$ &$6_5$ &$6_6$ & Polynomial& Cyclic?\\
\hline
17& [18,6]& [19,7]& [20,8]& [21,9]& $\fbox{[22,10]}$& [23,10]& 104B& No\\
18& [19,7]& [20,8]& [21,9]& [22,10]& $\fbox{[23,11]}$& [24,11]& 104B& No\\
19& [20,8]& [21,9]& [22,10]& [23,11]& [24,11]& $\fbox{[25,12]}$& 296D& No\\
20& [21,9]& [22,10]& [23,11]& [24,12]& [25,12]& $\fbox{[26,13]}$& 2041& No\\
21& [22,10]& [23,11]& [24,12]& $\fbox{[25,13]}$& [26,13]& [27,14]& 12CD& No\\
22& [23,11]& [24,12]& [25,13]& $\fbox{[26,14]}$& [27,14]& [28,15]& 1063& No\\
23& [24,12]& [25,13]& [26,14]& [27,15]& $\fbox{[28,16]}$& [29,16]& 1243& No\\
24& [25,13]& [26,14]& [27,15]& [28,16]& [29,17]& $\fbox{[30,18]}$& 1055& Yes\\
25& [26,14]& [27,15]& [28,16]& [29,17]& $\fbox{[30,18]}$& [31,18]& 1055& Yes\\
26& [27,15]& [28,16]& [29,17]& $\fbox{[30,18]}$& [31,18]& [32,18]& 1055& Yes\\
27& [28,16]& [29,17]& [30,18]& $\fbox{[31,19]}$& [32,19]& [33,19]& 1343& No\\
28& [29,17]& [30,18]& $\fbox{[31,19]}$& [32,19]& [33,20]& [34,20]& 1343& No\\
29& [30,18]& [31,19]& [32,20]& $\fbox{[33,21]}$& [34,21]& [35,22]& 187B& No\\
30& [31,19]& [32,20]& $\fbox{[33,21]}$& [34,21]& [35,22]& [36,22]& 187B& No\\
31& [32,20]& [33,21]& $\fbox{[34,22]}$& [35,22]& [36,23]& [37,23]& 1A7B& No\\
32& [33,21]& [34,22]& [35,22]& [36,23]& $\fbox{[37,24]}$& [38,24]& 2255& No\\
33& [34,22]& [35,22]& [36,23]& [37,24]& [38,25]& $\fbox{[39,26]}$& 2247& Yes\\
34& [35,22]& [36,23]& [37,24]& [38,25]& $\fbox{[39,26]}$& [40,26]& 2247& Yes\\
35& [36,23]& [37,24]& [38,25]& $\fbox{[39,26]}$& [40,26]& [41,27]& 20B5& No\\
36& [37,24]& [38,25]& [39,26]& [40,27]& $\fbox{[41,28]}$& [42,28]& 20B5& No\\
37& [38,25]& [39,26]& [40,27]& [41,28]& $\fbox{[42,29]}$& [43,29]& 24FF& No\\
38& [39,26]& [40,27]& [41,28]& [42,29]& $\fbox{[43,30]}$& [44,30]& 338B& No\\
39& [40,27]& [41,28]& [42,29]& $\fbox{[43,30]}$& [44,30]& [45,31]& 209D& No\\
40& [41,28]& [42,29]& [43,30]& [44,31]& $\fbox{[45,32]}$& [46,32]& 22F9& No\\
41& [42,29]& [43,30]& [44,31]& $\fbox{[45,32]}$& [46,32]& [47,33]& 21DD& No\\
42& [43,30]& [44,31]& [45,32]& $\fbox{[46,33]}$& [47,33]& [48,34]& 204F& No\\
43& [44,31]& [45,32]& [46,33]& $\fbox{[47,34]}$& [48,34]& [49,35]& 204F& No\\
44& [45,32]& [46,33]& [47,34]& [48,35]& $\fbox{[49,36]}$& [50,36]& 25C5& No\\
45& [46,33]& [47,34]& [48,35]& $\fbox{[49,36]}$& [50,36]& [51,36]& 2143& No\\
46& [47,34]& [48,35]& [49,36]& $\fbox{[50,37]}$& [51,37]& [52,38]& 20B5& No\\
47& [48,35]& [49,36]& [50,37]& $\fbox{[51,38]}$& [52,38]& [53,39]& 279B& No\\
48& [49,36]& [50,37]& [51,38]& $\fbox{[52,39]}$& [53,39]& [54,39]& 286D& No\\
49& [50,37]& [51,38]& [52,39]& [53,40]& $\fbox{[54,41]}$& [55,40]& 20B5& No\\
50& [51,38]& [52,39]& [53,40]& $\fbox{[54,41]}$& [55,41]& [56,41]& 204F& No\\
51& [52,39]& [53,40]& [54,41]& $\fbox{[55,42]}$& [56,42]& [57,43]& 20B5& No\\
52& [53,40]& [54,41]& [55,42]& $\fbox{[56,43]}$& [57,43]& [58,44]& 286D& No\\
53& [54,41]& [55,42]& $\fbox{[56,43]}$& [57,43]& [58,44]& [59,44]& 286D& No\\
54& [55,42]& [56,43]& [57,44]& $\fbox{[58,45]}$& [59,45]& [60,45]& 2BCF& No\\
55& [56,43]& [57,44]& $\fbox{[58,45]}$& [59,45]& [60,46]& [61,47]& 24FF& No\\
56& [57,44]& [58,45]& [59,46]& $\fbox{[60,47]}$& [61,47]& [62,48]& 24FF& No\\
57& [58,45]& [59,46]& [60,47]& [61,48]& [62,49]& $\fbox{[63,50]}$& 24FF& Yes\\
58& [59,46]& [60,47]& [61,48]& [62,49]& $\fbox{[63,50]}$& [64,50]& 24FF& Yes\\
59& [60,47]& [61,48]& [62,49]& $\fbox{[63,50]}$& [64,50]& [65,50]& 24FF& Yes\\
60& [61,48]& [62,49]& [63,50]& $\fbox{[64,51]}$& [65,51]& [66,51]& 29CB& No\\
\hline
\end{tabular}
}
\end{center}
\begin{center}
\caption{Optimal (shortened) cyclic codes correcting bursts of length up
to 6, for a guard\newline space from $g = 17$ to $g = 60$}
\end{center}
\end{table}

\begin{table}
\begin{center}
\tiny{
\begin{tabular}{|c|c|c|c|c|c|c|c|c|}
\hline
$g$ &$6_1$ &$6_2$ &$6_3$ &$6_4$ &$6_5$ &$6_6$ &Polynomial&Cyclic?\\
\hline
61& [62,49]& [63,50]& $\fbox{[64,51]}$& [65,51]& [66,52]& [67,52]& 21CB& No\\
62& [63,50]& [64,51]& $\fbox{[65,52]}$& [66,52]& [67,53]& [68,53]& 2BCF& No\\
63& [64,51]& [65,52]& $\fbox{[66,53]}$& [67,53]& [68,54]& [69,54]& 2BCF& No\\
64& [65,52]& $\fbox{[66,53]}$& [67,53]& [68,54]& [69,55]& [70,55]& 28DB& No\\
65& [66,53]& $\fbox{[67,54]}$& [68,54]& [69,55]& [70,56]& [71,56]& 29CB& No\\
66& $\fbox{[67,54]}$& [68,54]& [69,55]& [70,56]& [71,57]& [72,57]& 29CB& No\\
67& [68,54]& [69,55]& [70,56]& [71,57]& $\fbox{[72,58]}$& [73,58]& 41AD& No\\
68& [69,55]& [70,56]& [71,57]& [72,58]& $\fbox{[73,59]}$& [74,59]& 44D1& No\\
69& [70,56]& [71,57]& [72,58]& [73,59]& $\fbox{[74,60]}$& [75,60]& 6897& No\\
70& [71,57]& [72,58]& [73,59]& [74,60]& $\fbox{[75,61]}$& [76,61]& 708F& No\\
71& [72,58]& [73,59]& [74,60]& [75,61]& $\fbox{[76,62]}$& [77,62]& 56F7& No\\
72& [73,59]& [74,60]& [75,61]& [76,62]& $\fbox{[77,63]}$& [78,63]& 6E6F& No\\
73& [74,60]& [75,61]& [76,62]& [77,63]& $\fbox{[78,64]}$& [79,64]& 6E6F& No\\
74& [75,61]& [76,62]& [77,63]& $\fbox{[78,64]}$& [79,64]& [80,65]& 410B& No\\
75& [76,62]& [77,63]& [78,64]& $\fbox{[79,65]}$& [80,65]& [81,66]& 4125& No\\
76& [77,63]& [78,64]& [79,65]& [80,66]& $\fbox{[81,67]}$& [82,67]& 4863& No\\
77& [78,64]& [79,65]& [80,66]& [81,67]& $\fbox{[82,68]}$& [83,68]& 708F& No\\
78& [79,65]& [80,66]& [81,67]& [82,68]& $\fbox{[83,69]}$& [84,69]& 46F9& No\\
79& [80,66]& [81,67]& [82,68]& [83,69]& $\fbox{[84,70]}$& [85,70]& 77CF& No\\
80& [81,67]& [82,68]& [83,69]& [84,70]& [85,71]& $\fbox{[86,72]}$& 406D& No\\
81& [82,68]& [83,69]& [84,70]& [85,71]& [86,72]& $\fbox{[87,73]}$& 406D& No\\
82& [83,69]& [84,70]& [85,71]& [86,72]& $\fbox{[87,73]}$& [88,73]& 406D& No\\
83& [84,70]& [85,71]& [86,72]& [87,73]& $\fbox{[88,74]}$& [89,74]& 40BD& No\\
84& [85,71]& [86,72]& [87,73]& $\fbox{[88,74]}$& [89,74]& [90,75]& 40B1& No\\
85& [86,72]& [87,73]& [88,74]& $\fbox{[89,75]}$& [90,75]& [91,76]& 43B5& No\\
86& [87,73]& [88,74]& [89,75]& [90,76]& $\fbox{[91,77]}$& [92,77]& 49C7& No\\
87& [88,74]& [89,75]& [90,76]& $\fbox{[91,77]}$& [92,77]& [93,78]& 43B5& No\\
88& [89,75]& [90,76]& [91,77]& $\fbox{[92,78]}$& [93,78]& [94,79]& 414F& No\\
89& [90,76]& [91,77]& [92,78]& $\fbox{[93,79]}$& [94,79]& [95,79]& 40B1& No\\
90& [91,77]& [92,78]& [93,79]& $\fbox{[94,80]}$& [95,80]& [96,80]& 40B1& No\\
91& [92,78]& [93,79]& [94,80]& $\fbox{[95,81]}$& [96,81]& [97,82]& 49C7& No\\
92& [93,79]& [94,80]& [95,81]& $\fbox{[96,82]}$& [97,82]& [98,83]& 44D1& No\\
93& [94,80]& [95,81]& [96,82]& $\fbox{[97,83]}$& [98,83]& [99,84]& 4547& No\\
94& [95,81]& [96,82]& [97,83]& $\fbox{[98,84]}$& [99,84]& [100,85]& 4251& No\\
95& [96,82]& [97,83]& [98,84]& $\fbox{[99,85]}$& [100,85]& [101,85]& 47BD& No\\
96& [97,83]& [98,84]& [99,85]& $\fbox{[100,86]}$& [101,86]& [102,87]& 430F& No\\
97& [98,84]& [99,85]& [100,86]& $\fbox{[101,87]}$& [102,87]& [103,88]& 4125& No\\
98& [99,85]& [100,86]& [101,87]& $\fbox{[102,88]}$& [103,88]& [104,89]& 40B1& No\\
99& [10086]& [10187]& [10288]& [10389]& [10490]& $\fbox{[10591]}$& 42BF& Yes\\
100& [101,87]& [102,88]& [103,89]& [104,90]& $\fbox{[105,91]}$& [106,91]& 42BF& Yes\\
\hline
\end{tabular}
}
\end{center}
\begin{center}
\caption{Optimal (shortened) cyclic codes correcting bursts of length up
to 6, for a guard\newline space from $g = 61$ to $g = 100$}
\end{center}
\end{table}

\begin{table}
\begin{center}
\tiny{
\begin{tabular}{|c|c|c|c|c|c|c|c|c|c|}
\hline
$g$ &$7_1$ &$7_2$ &$7_3$ &$7_4$ &$7_5$ &$7_6$ &$7_7$& Polynomial& Cyclic?\\
\hline
20& [21,7]& [22,8]& [23,9]& [24,10]& [25,11]& $\fbox{[26,12]}$& [27,12]& 40C3& No\\
21& [22,8]& [23,9]& [24,10]& [25,11]& [26,12]& $\fbox{[27,13]}$& [28,13]& 42F3& No\\
22& [23,9]& [24,10]& [25,11]& [26,12]& [27,13]& [28,13]& $\fbox{[29,14]}$& E9CF& No\\
23& [24,10]& [25,11]& [26,12]& [27,13]& [28,14]& [29,14]& $\fbox{[30,15]}$& 8081& No\\
24& [25,11]& [26,12]& [27,13]& [28,14]& $\fbox{[29,15]}$& [30,15]& [31,16]& 40DB& No\\
25& [26,12]& [27,13]& [28,14]& [29,15]& $\fbox{[30,16]}$& [31,16]& [32,16]& 40F5& No\\
26& [27,13]& [28,14]& [29,15]& [30,16]& $\fbox{[31,17]}$& [32,17]& [33,17]& 5CAF& No\\
27& [28,14]& [29,15]& [30,16]& [31,17]& $\fbox{[32,18]}$& [33,18]& [34,18]& 559F& No\\
28& [29,15]& [30,16]& [31,17]& [32,18]& [33,18]& [34,19]& $\fbox{[35,20]}$& CEAF& Yes\\
29& [30,16]& [31,17]& [32,18]& [33,19]& [34,19]& [35,20]& $\fbox{[36,21]}$& B4FD& No\\
30& [31,17]& [32,18]& [33,19]& $\fbox{[34,20]}$& [35,20]& [36,21]& [37,21]& 40B9& No\\
31& [32,18]& [33,19]& [34,20]& $\fbox{[35,21]}$& [36,21]& [37,22]& [38,22]& 40B9& No\\
32& [33,19]& [34,20]& [35,21]& [36,21]& [37,22]& $\fbox{[38,23]}$& [39,23]& 8171& No\\
33& [34,20]& [35,21]& [36,22]& [37,22]& [38,23]& $\fbox{[39,24]}$& [40,24]& 84DF& No\\
34& [35,21]& [36,22]& [37,23]& [38,23]& [39,24]& $\fbox{[40,25]}$& [41,25]& 80FB& No\\
35& [36,22]& [37,23]& [38,23]& [39,24]& [40,25]& $\fbox{[41,26]}$& [42,26]& A195& No\\
36& [37,23]& [38,24]& [39,24]& [40,25]& [41,26]& $\fbox{[42,27]}$& [43,27]& 80BD& No\\
37& [38,24]& [39,24]& [40,25]& [41,26]& [42,27]& $\fbox{[43,28]}$& [44,28]& 80CD& No\\
38& [39,24]& [40,25]& [41,26]& [42,27]& $\fbox{[43,28]}$& [44,28]& [45,29]& 80CD& No\\
39& [40,25]& [41,26]& [42,27]& [43,28]& [44,29]& $\fbox{[45,30]}$& [46,30]& 8171& No\\
40& [41,26]& [42,27]& [43,28]& [44,29]& $\fbox{[45,30]}$& [46,30]& [47,31]& 8171& No\\
41& [42,27]& [43,28]& [44,29]& [45,30]& [46,31]& $\fbox{[47,32]}$& [48,32]& A10D& No\\
42& [43,28]& [44,29]& [45,30]& [46,31]& [47,32]& $\fbox{[48,33]}$& [49,33]& B757& No\\
43& [44,29]& [45,30]& [46,31]& [47,32]& [48,33]& $\fbox{[49,34]}$& [50,34]& 98F7& No\\
44& [45,30]& [46,31]& [47,32]& [48,33]& $\fbox{[49,34]}$& [50,34]& [51,35]& 80CB& No\\
45& [46,31]& [47,32]& [48,33]& [49,34]& $\fbox{[50,35]}$& [51,35]& [52,36]& 97B7& No\\
46& [47,32]& [48,33]& [49,34]& [50,35]& $\fbox{[51,36]}$& [52,36]& [53,36]& 80CB& No\\
47& [48,33]& [49,34]& [50,35]& [51,36]& $\fbox{[52,37]}$& [53,37]& [54,38]& 8B1F& No\\
48& [49,34]& [50,35]& [51,36]& [52,37]& $\fbox{[53,38]}$& [54,38]& [55,39]& 89F1& No\\
49& [50,35]& [51,36]& [52,37]& [53,38]& $\fbox{[54,39]}$& [55,39]& [56,40]& AF4D& No\\
50& [51,36]& [52,37]& [53,38]& [54,39]& $\fbox{[55,40]}$& [56,40]& [57,41]& 8597& No\\
51& [52,37]& [53,38]& [54,39]& [55,40]& $\fbox{[56,41]}$& [57,41]& [58,42]& 9B43& No\\
52& [53,38]& [54,39]& [55,40]& $\fbox{[56,41]}$& [57,41]& [58,42]& [59,42]& 812D& No\\
53& [54,39]& [55,40]& [56,41]& [57,42]& $\fbox{[58,43]}$& [59,43]& [60,43]& 878F& No\\
54& [55,40]& [56,41]& [57,42]& $\fbox{[58,43]}$& [59,43]& [60,44]& [61,44]& 878F& No\\
55& [56,41]& [57,42]& [58,43]& [59,44]& $\fbox{[60,45]}$& [61,45]& [62,46]& B39B& No\\
56& [57,42]& [58,43]& [59,44]& [60,45]& [61,46]& [62,47]& $\fbox{[63,48]}$& 8B1F& Yes\\
57& [58,43]& [59,44]& [60,45]& [61,46]& [62,47]& $\fbox{[63,48]}$& [64,48]& 8B1F& Yes\\
58& [59,44]& [60,45]& [61,46]& [62,47]& $\fbox{[63,48]}$& [64,48]& [65,49]& 878F& No\\
59& [60,45]& [61,46]& [62,47]& $\fbox{[63,48]}$& [64,48]& [65,49]& [66,49]& 878F& No\\
60& [61,46]& [62,47]& [63,48]& $\fbox{[64,49]}$& [65,49]& [66,50]& [67,50]& 9D9D& No\\
\hline
\end{tabular}
}
\end{center}
\begin{center}
\caption{Optimal (shortened) cyclic codes correcting bursts of length up
to 7, for a guard\newline space from $g = 20$ to $g = 60$}
\end{center}
\end{table}

\begin{table}
\begin{center}
\tiny{
\begin{tabular}{|c|c|c|c|c|c|c|c|c|c|}
\hline
$g$ &$7_1$ &$7_2$ &$7_3$ &$7_4$ &$7_5$ &$7_6$ &$7_7$& Polynomial& Cyclic?\\
\hline
61& [62,47]& [63,48]& [64,49]& [65,50]& $\fbox{[66,51]}$& [67,51]& [68,51]& C8E7& No\\
62& [63,48]& [64,49]& [65,50]& [66,51]& $\fbox{[67,52]}$& [68,52]& [69,52]& 91CD& No\\
63& [64,49]& [65,50]& [66,51]& $\fbox{[67,52]}$& [68,52]& [69,53]& [70,53]& 91CD& No\\
64& [65,50]& [66,51]& [67,52]& $\fbox{[68,53]}$& [69,53]& [70,54]& [71,54]& 91CD& No\\
65& [66,51]& [67,52]& [68,53]& [69,54]& $\fbox{[70,55]}$& [71,55]& [72,55]& 8F19& No\\
66& [67,52]& [68,53]& [69,54]& $\fbox{[70,55]}$& [71,55]& [72,56]& [73,56]& 89F1& No\\
67& [68,53]& [69,54]& [70,55]& $\fbox{[71,56]}$& [72,56]& [73,57]& [74,57]& 8F19& No\\
68& [69,54]& [70,55]& [71,56]& $\fbox{[72,57]}$& [73,57]& [74,58]& [75,58]& 89F1& No\\
69& [70,55]& [71,56]& [72,57]& [73,58]& $\fbox{[74,59]}$& [75,59]& [76,59]& 9D9D& No\\
70& [71,56]& [72,57]& [73,58]& $\fbox{[74,59]}$& [75,59]& [76,60]& [77,60]& 91CD& No\\
71& [72,57]& [73,58]& [74,59]& $\fbox{[75,60]}$& [76,60]& [77,61]& [78,61]& 91CD& No\\
72& [73,58]& [74,59]& $\fbox{[75,60]}$& [76,60]& [77,61]& [78,62]& [79,62]& 91CD& No\\
73& [74,59]& $\fbox{[75,60]}$& [76,60]& [77,61]& [78,62]& [79,63]& [80,63]& 89F1& No\\
74& [75,60]& [76,61]& [77,62]& $\fbox{[78,63]}$& [79,63]& [80,64]& [81,64]& 89F1& No\\
75& [76,61]& [77,62]& [78,63]& $\fbox{[79,64]}$& [80,64]& [81,65]& [82,65]& 8F19& No\\
76& [77,62]& [78,63]& $\fbox{[79,64]}$& [80,64]& [81,65]& [82,66]& [83,66]& 8F19& No\\
77& [78,63]& [79,64]& $\fbox{[80,65]}$& [81,65]& [82,66]& [83,66]& [84,67]& 89F1& No\\
78& [79,64]& [80,65]& [81,66]& $\fbox{[82,67]}$& [83,67]& [84,68]& [85,69]& 8F19& No\\
79& [80,65]& [81,66]& [82,67]& $\fbox{[83,68]}$& [84,68]& [85,69]& [86,69]& 8F19& No\\
80& [81,66]& [82,67]& $\fbox{[83,68]}$& [84,68]& [85,69]& [86,70]& [87,70]& 8F19& No\\
81& [82,67]& [83,68]& [84,69]& $\fbox{[85,70]}$& [86,70]& [87,70]& [88,71]&
8F19& No\\
82& [83,68]& [84,69]& [85,70]& $\fbox{[86,71]}$& [87,71]& [88,72]& [89,72]& 9D9D& No\\
83& [84,69]& [85,70]& $\fbox{[86,71]}$& [87,71]& [88,72]& [89,73]& [90,73]& 8F19& No\\
84& [85,70]& $\fbox{[86,71]}$& [87,71]& [88,72]& [89,73]& [90,74]& [91,74]& 8F19& No\\
85& [86,71]& $\fbox{[87,72]}$& [88,72]& [89,73]& [90,74]& [91,74]& [92,75]& 9D9D& No\\
86& [87,72]& [88,73]& $\fbox{[89,74]}$& [90,74]& [91,75]& [92,76]& [93,77]& 8F19& No\\
87& [88,73]& $\fbox{[89,74]}$& [90,74]& [91,75]& [92,76]& [93,77]& [94,77]& 8F19& No\\
88& [89,74]& [90,75]& $\fbox{[91,76]}$& [92,76]& [93,77]& [94,77]& [95,78]& 8F19& No\\
89& [90,75]& $\fbox{[91,76]}$& [92,76]& [93,77]& [94,78]& [95,78]& [96,79]& 8F19& No\\
90& $\fbox{[91,76]}$& [92,76]& [93,77]& [94,78]& [95,79]& [96,79]& [97,80]& 8F19& No\\
91& $\fbox{[92,77]}$& [93,77]& [94,78]& [95,79]& [96,80]& [97,80]& [98,81]& 8F19& No\\
92& [93,78]& [94,79]& $\fbox{[95,80]}$& [96,80]& [97,81]& [98,81]& [99,82]& 8F19& No\\
93& [94,79]& $\fbox{[95,80]}$& [96,80]& [97,81]& [98,82]& [99,83]& [100,82]& 8F19& No\\
94& $\fbox{[95,80]}$& [96,80]& [97,81]& [98,82]& [99,83]& [100,83]& [101,84]& 8F19& No\\
95& [96,81]& [97,82]& $\fbox{[98,83]}$& [99,83]& [100,84]& [101,85]& [102,86]& 8F19& No\\
96& [97,82]& $\fbox{[98,83]}$& [99,83]& [100,84]& [101,85]& [102,86]& [103,86]& 8F19& No\\
97& $\fbox{[98,83]}$& [99,83]& [100,84]& [101,85]& [102,86]& [103,86]& [104,86]& 8F19& No\\
98& [99,84]& [100,85]& [101,86]& $\fbox{[102,87]}$& [103,87]& [104,88]& [105,89]& 8F19& No\\
99& [100,85]& [101,86]& $\fbox{[102,87]}$& [103,87]& [104,88]& [105,89]& [106,89]& 8F19& No\\
100& [101,86]& $\fbox{[102,87]}$& [103,87]& [104,88]& [105,89]& [106,89]& [107,90]& 8F19& No\\
\hline
\end{tabular}
}
\end{center}
\begin{center}
\caption{Optimal (shortened) cyclic codes correcting bursts of length up
to 7, for a guard\newline space from $g = 61$ to $g = 100$}
\end{center}
\end{table}

\begin{table}
\begin{center}
\tiny{
\begin{tabular}{|c|c|c|c|c|c|c|c|c|c|c|}
\hline
$g$ &$8_1$ &$8_2$ &$8_3$ &$8_4$ &$8_5$ &$8_6$ &$8_7$ &$8_8$&Polynomial&Cyclic?\\
\hline
20& [21,5]& [22,6]& [23,7]& [24,8]& [25,9]& [26,10]& [27,11]& $\fbox{[28,12]}$& 10111&Yes\\
21& [22,6]& [23,7]& [24,8]& [25,9]& [26,10]& [27,11]& $\fbox{[28,12]}$& [29,12]& 10111&Yes\\
22& [23,7]& [24,8]& [25,9]& [26,10]& [27,11]& [28,12]& [29,13]& $\fbox{[30,14]}$& 10115&Yes\\
23& [24,8]& [25,9]& [26,10]& [27,11]& [28,12]& [29,13]& [30,14]& $\fbox{[31,15]}$& 10117&Yes\\
24& [25,9]& [26,10]& [27,11]& [28,12]& [29,13]& [30,14]& $\fbox{[31,15]}$& [32,15]& 10117&Yes\\
25& [26,10]& [27,11]& [28,12]& [29,13]& [30,14]& [31,15]& $\fbox{[32,16]}$& [33,16]& 19F17&No\\
26& [27,11]& [28,12]& [29,13]& [30,14]& [31,15]& [32,16]& $\fbox{[33,17]}$& [34,17]& 15B2D&No\\
27& [28,12]& [29,13]& [30,14]& [31,15]& [32,16]& [33,17]& [34,18]& $\fbox{[35,19]}$& 15533&No\\
28& [29,13]& [30,14]& [31,15]& [32,16]& [33,17]& [34,18]& $\fbox{[35,19]}$& [36,19]& 15533&No\\
29& [30,14]& [31,15]& [32,16]& [33,17]& [34,18]& $\fbox{[35,19]}$& [36,19]& [37,19]& 11109&No\\
30& [31,15]& [32,16]& [33,17]& [34,18]& [35,19]& [36,20]& [37,21]& $\fbox{[38,22]}$& 11105&No\\
31& [32,16]& [33,17]& [34,18]& [35,19]& [36,20]& [37,21]& $\fbox{[38,22]}$& [39,22]& 11105&No\\
32& [33,17]& [34,18]& [35,19]& [36,20]& [37,21]& $\fbox{[38,22]}$& [39,22]& [40,23]& 11105&No\\
33& [34,18]& [35,19]& [36,20]& [37,21]& [38,22]& $\fbox{[39,23]}$& [40,23]& [41,24]& 1B1B7&No\\
34& [35,19]& [36,20]& [37,21]& [38,22]& $\fbox{[39,23]}$& [40,23]& [41,24]& [42,24]& 1B1B7&No\\
35& [36,20]& [37,21]& [38,22]& [39,23]& $\fbox{[40,24]}$& [41,24]& [42,25]& [43,25]& 1B1B7&No\\
36& [37,21]& [38,22]& [39,23]& [40,24]& $\fbox{[41,25]}$& [42,25]& [43,26]& [44,26]& 15103&No\\
37& [38,22]& [39,23]& [40,24]& [41,25]& [42,25]& [43,26]& [44,27]& $\fbox{[45,28]}$& 2B173&Yes\\
38& [39,23]& [40,24]& [41,25]& [42,25]& [43,26]& [44,27]& $\fbox{[45,28]}$& [46,28]& 213D1&No\\
39& [40,24]& [41,25]& [42,26]& [43,27]& $\fbox{[44,28]}$& [45,28]& [46,29]& [47,29]& 11953&No\\
40& [41,25]& [42,26]& [43,27]& [44,28]& [45,28]& [46,29]& $\fbox{[47,30]}$& [48,30]& 29EDF&No\\
41& [42,26]& [43,27]& [44,28]& [45,29]& $\fbox{[46,30]}$& [47,30]& [48,31]& [49,31]& 11953&No\\
42& [43,27]& [44,28]& [45,29]& $\fbox{[46,30]}$& [47,30]& [48,31]& [49,31]& [50,32]& 11953&No\\
43& [44,28]& [45,29]& [46,30]& [47,31]& [48,32]& [49,32]& [50,33]& $\fbox{[51,34]}$& 299FB&Yes\\
44& [45,29]& [46,30]& [47,31]& [48,32]& [49,32]& [50,33]& [51,34]& $\fbox{[52,35]}$& 241A9&No\\
45& [46,30]& [47,31]& [48,32]& [49,33]& $\fbox{[50,34]}$& [51,34]& [52,35]& [53,35]& 12959&No\\
46& [47,31]& [48,32]& [49,33]& $\fbox{[50,34]}$& [51,34]& [52,35]& [53,35]& [54,36]& 11953&No\\
47& [48,32]& [49,33]& [50,34]& [51,34]& [52,35]& [53,36]& $\fbox{[54,37]}$& [55,37]& 205BF&No\\
48& [49,33]& [50,34]& [51,34]& [52,35]& [53,36]& [54,37]& $\fbox{[55,38]}$& [56,38]& 26177&No\\
49& [50,34]& [51,34]& [52,35]& [53,36]& [54,37]& $\fbox{[55,38]}$& [56,38]& [57,39]& 201B5&No\\
50& [51,34]& [52,35]& [53,36]& [54,37]& [55,38]& $\fbox{[56,39]}$& [57,39]& [58,39]& 20105&No\\
51& [52,35]& [53,36]& [54,37]& [55,38]& [56,39]& $\fbox{[57,40]}$& [58,40]& [59,40]& 21A4B&No\\
52& [53,36]& [54,37]& [55,38]& [56,39]& [57,40]& [58,41]& $\fbox{[59,42]}$& [60,41]& 28A1D&No\\
53& [54,37]& [55,38]& [56,39]& [57,40]& [58,41]& $\fbox{[59,42]}$& [60,42]& [61,43]& 201AD&No\\
54& [55,38]& [56,39]& [57,40]& [58,41]& [59,42]& $\fbox{[60,43]}$& [61,43]& [62,44]& 202AB&No\\
55& [56,39]& [57,40]& [58,41]& [59,42]& [60,43]& [61,44]& [62,45]& $\fbox{[63,46]}$& 2EA37&Yes\\
56& [57,40]& [58,41]& [59,42]& [60,43]& [61,44]& [62,45]& $\fbox{[63,46]}$& [64,46]& 2EA37&Yes\\
57& [58,41]& [59,42]& [60,43]& [61,44]& [62,45]& $\fbox{[63,46]}$& [64,46]& [65,46]& 20267&No\\
58& [59,42]& [60,43]& [61,44]& [62,45]& [63,46]& $\fbox{[64,47]}$& [65,47]& [66,47]& 213C5&No\\
59& [60,43]& [61,44]& [62,45]& [63,46]& [64,47]& $\fbox{[65,48]}$& [66,48]& [67,48]& 2423F&No\\
60& [61,44]& [62,45]& [63,46]& [64,47]& [65,48]& $\fbox{[66,49]}$& [67,49]& [68,49]& 32D27&No\\
\hline
\end{tabular}
}
\end{center}
\begin{center}
\caption{Optimal (shortened) cyclic codes correcting bursts of length up
to 8, for a guard\newline space from $g = 20$ to $g = 60$}
\end{center}
\end{table}

\begin{table}
\begin{center}
\tiny{
\begin{tabular}{|c|c|c|c|c|c|c|c|c|c|c|}
\hline
$g$ &$8_1$ &$8_2$ &$8_3$ &$8_4$ &$8_5$ &$8_6$ &$8_7$ &$8_8$&Polynomial&Cyclic?\\
\hline
61& [62,45]& [63,46]& [64,47]& [65,48]& $\fbox{[66,49]}$& [67,49]& [68,50]& [69,50]& 20171&No\\
62& [63,46]& [64,47]& [65,48]& [66,49]& [67,50]& $\fbox{[68,51]}$& [69,51]& [70,51]& 221FD&No\\
63& [64,47]& [65,48]& [66,49]& [67,50]& [68,51]& $\fbox{[69,52]}$& [70,52]& [71,52]& 2838B&No\\
64& [65,48]& [66,49]& [67,50]& [68,51]& $\fbox{[69,52]}$& [70,52]& [71,53]& [72,54]& 2838B&No\\
65& [66,49]& [67,50]& [68,51]& [69,52]& $\fbox{[70,53]}$& [71,53]& [72,54]& [73,55]& 20A07&No\\
66& [67,50]& [68,51]& [69,52]& [70,53]& $\fbox{[71,54]}$& [72,54]& [73,55]& [74,55]& 209FB&No\\
67& [68,51]& [69,52]& [70,53]& [71,54]& [72,55]& $\fbox{[73,56]}$& [74,56]& [75,56]& 355DF&No\\
68& [69,52]& [70,53]& [71,54]& [72,55]& $\fbox{[73,56]}$& [74,56]& [75,57]& [76,57]& 21217&No\\
69& [70,53]& [71,54]& [72,55]& [73,56]& $\fbox{[74,57]}$& [75,57]& [76,58]& [77,58]& 2EE23&No\\
70& [71,54]& [72,55]& [73,56]& $\fbox{[74,57]}$& [75,58]& [76,58]& [77,59]& [78,61]& 21217&No\\
71& [72,55]& [73,56]& [74,57]& [75,58]& $\fbox{[76,59]}$& [77,59]& [78,60]& [79,60]& 20A07&No\\
72& [73,56]& [74,57]& [75,58]& [76,59]& $\fbox{[77,60]}$& [78,60]& [79,61]& [80,61]& 25597&No\\
73& [74,58]& [75,58]& [76,59]& [77,60]& $\fbox{[78,61]}$& [79,61]& [80,62]& [81,62]& 33DDB&No\\
74& [75,58]& [76,59]& [77,60]& [78,61]& $\fbox{[79,62]}$& [80,62]& [81,63]& [82,63]& 223B7&No\\
75& [76,59]& [77,60]& [78,61]& [79,62]& $\fbox{[80,63]}$& [81,63]& [82,64]& [83,64]& 21217&No\\
76& [77,60]& [78,61]& [79,62]& [80,63]& $\fbox{[81,64]}$& [82,64]& [83,65]& [84,65]& 223B7&No\\
77& [78,61]& [79,62]& [80,63]& [81,64]& [82,65]& [83,66]& [84,67]& $\fbox{[85,68]}$& 3B68F&Yes\\
78& [79,62]& [80,63]& [81,64]& [82,65]& [83,66]& [84,67]& $\fbox{[85,68]}$& [86,68]& 3B68F&Yes\\
79& [80,63]& [81,64]& [82,65]& [83,66]& [84,67]& $\fbox{[85,68]}$& [86,68]& [87,69]& 3B68F&Yes\\
80& [81,64]& [82,65]& [83,66]& [84,67]& $\fbox{[85,68]}$& [86,68]& [87,69]& [88,69]& 3B68F&Yes\\
81& [82,65]& [83,66]& [84,67]& $\fbox{[85,68]}$& [86,68]& [87,69]& [88,70]& [89,70]& 33DDB&No\\
82& [83,66]& [84,67]& [85,68]& $\fbox{[86,69]}$& [87,69]& [88,70]& [89,71]& [90,71]& 355DF&No\\
83& [84,67]& [85,68]& [86,69]& [87,70]& $\fbox{[88,71]}$& [89,71]& [90,72]& [91,73]& 21217&No\\
84& [85,68]& [86,69]& [87,70]& $\fbox{[88,71]}$& [89,71]& [90,72]& [91,73]& [92,73]& 21217&No\\
85& [86,69]& [87,70]& [88,71]& $\fbox{[89,72]}$& [90,72]& [91,73]& [92,74]& [93,75]& 20105&No\\
86& [87,70]& [88,71]& [89,72]& [90,73]& $\fbox{[91,74]}$& [92,74]& [93,75]& [94,75]& 20105&No\\
87& [88,71]& [89,72]& [90,73]& $\fbox{[91,74]}$& [92,74]& [93,75]& [94,76]& [95,76]& 20105&No\\
88& [89,72]& [90,73]& [91,74]& $\fbox{[92,75]}$& [93,75]& [94,76]& [95,77]& [96,77]& 20105&No\\
89& [90,73]& [91,74]& $\fbox{[92,75]}$& [93,75]& [94,76]& [95,77]& [96,77]& [97,78]& 20105&No\\
90& [91,74]& [92,75]& [93,76]& $\fbox{[94,77]}$& [95,77]& [96,78]& [97,79]& [98,79]& 20105&No\\
91& [92,75]& [93,76]& [94,77]& $\fbox{[95,78]}$& [96,78]& [97,79]& [98,80]& [99,80]& 20105&No\\
92& [93,76]& [94,77]& $\fbox{[95,78]}$& [96,78]& [97,79]& [98,80]& [99,80]& [100,81]& 20105&No\\
93& [94,77]& $\fbox{[95,78]}$& [96,78]& [97,79]& [98,80]& [99,81]& [100,81]& [101,82]& 20105&No\\
94& [95,78]& [96,78]& [97,79]& [98,80]& [99,81]& [100,82]& [101,83]& $\fbox{[102,84]}$& 6DB07&Yes\\
95& [96,79]& [97,79]& [98,80]& [99,81]& [100,82]& [101,83]& $\fbox{[102,84]}$& [103,84]& 6DB07&Yes\\
96& [97,79]& [98,80]& [99,81]& [100,82]& [101,83]& $\fbox{[102,84]}$& [103,84]& [104,85]& 40951&No\\
97& [98,80]& [99,81]& [100,82]& [101,83]& [102,84]& [103,85]& [104,86]& $\fbox{[105,87]}$& 461B9&Yes\\
98& [99,81]& [100,82]& [101,83]& [102,84]& [103,85]& [104,86]& $\fbox{[105,87]}$& [106,87]& 461B9&Yes\\
99& [100,82]& [101,83]& [102,84]& [103,85]& [104,86]& $\fbox{[105,87]}$& [106,87]& [107,88]& 40D29&No\\
100& [101,83]& [102,84]& [103,85]& [104,86]& [105,87]& $\fbox{[106,88]}$& [107,88]& [108,89]& 402C9&No\\
\hline
\end{tabular}
}
\end{center}
\begin{center}
\caption{Optimal (shortened) cyclic codes correcting bursts of length up
to 8, for a guard\newline space from $g = 61$ to $g = 100$}
\end{center}
\end{table}

\begin{table}
\begin{center}
\tiny{
\begin{tabular}{|c|c|c|c|c|c|c|c|c|c|c|c|}
\hline
$g$ &$9_1$ &$9_2$ &$9_3$ &$9_4$ &$9_5$ &$9_6$ &$9_7$ &$9_8$ &$9_9$&Polynomial&Cyclical?\\
\hline
20& [21,3]& [22,4]& [23,5]& [24,6]& [25,7]& [26,8]& [27,9]& $\fbox{[28,10]}$& [29,10]& 4936D&No\\
21& [22,4]& [23,5]& [24,6]& [25,7]& [26,8]& [27,9]& [28,10]& [29,11]& $\fbox{[30,12]}$& 50325&No\\
22& [23,5]& [24,6]& [25,7]& [26,8]& [27,9]& [28,10]& [29,11]& $\fbox{[30,12]}$& [31,12]& 40245&No\\
23& [24,6]& [25,7]& [26,8]& [27,9]& [28,10]& [29,11]& [30,12]& [31,12]& $\fbox{[32,13]}$& 862F7&No\\
24& [25,7]& [26,8]& [27,9]& [28,10]& [29,11]& [30,12]& [31,13]& $\fbox{[32,14]}$& [33,14]& 51E43&No\\
25& [26,8]& [27,9]& [28,10]& [29,11]& [30,12]& [31,13]& [32,14]& $\fbox{[33,15]}$& [34,15]& 51E43&No\\
26& [27,9]& [28,10]& [29,11]& [30,12]& [31,13]& [32,14]& [33,15]& $\fbox{[34,16]}$& [35,16]& 40303&No\\
27& [28,10]& [29,11]& [30,12]& [31,13]& [32,14]& [33,15]& [34,16]& [35,16]& $\fbox{[36,17]}$& 9F5BB&No\\
28& [29,11]& [30,12]& [31,13]& [32,14]& [33,15]& [34,16]& [35,17]& [36,17]& $\fbox{[37,18]}$& 9AB45&No\\
29& [30,12]& [31,13]& [32,14]& [33,15]& [34,16]& [35,17]& [36,18]& [37,18]& $\fbox{[38,19]}$& 80201&No\\
30& [31,13]& [32,14]& [33,15]& [34,16]& [35,17]& [36,18]& $\fbox{[37,19]}$& [38,19]& [39,19]& 45BF9&No\\
31& [32,14]& [33,15]& [34,16]& [35,17]& [36,18]& [37,19]& $\fbox{[38,20]}$& [39,20]& [40,20]& 403C9&No\\
32& [33,15]& [34,16]& [35,17]& [36,18]& [37,19]& $\fbox{[38,20]}$& [39,20]& [40,21]& [41,21]& 403C9&No\\
33& [34,16]& [35,17]& [36,18]& [37,19]& [38,20]& $\fbox{[39,21]}$& [40,21]& [41,22]& [42,22]& 42633&No\\
34& [35,17]& [36,18]& [37,19]& [38,20]& [39,21]& [40,22]& $\fbox{[41,23]}$& [42,23]& [43,23]& 5EEDB&No\\
35& [36,18]& [37,19]& [38,20]& [39,21]& [40,22]& [41,23]& $\fbox{[42,24]}$& [43,24]& [44,24]& 50BFF&No\\
36& [37,19]& [38,20]& [39,21]& [40,22]& [41,23]& [42,24]& [43,25]& [44,26]& $\fbox{[45,27]}$& 40249&Yes\\
37& [38,20]& [39,21]& [40,22]& [41,23]& [42,24]& [43,25]& [44,26]& $\fbox{[45,27]}$& [46,26]& 40249&Yes\\
38& [39,21]& [40,22]& [41,23]& [42,24]& [43,25]& [44,26]& $\fbox{[45,27]}$& [46,27]& [47,27]& 40249&Yes\\
39& [40,22]& [41,23]& [42,24]& [43,25]& [44,26]& $\fbox{[45,27]}$& [46,27]& [47,28]& [48,28]& 40249&Yes\\
40& [41,23]& [42,24]& [43,25]& [44,26]& [45,27]& $\fbox{[46,28]}$& [47,28]& [48,29]& [49,29]& 4362F&No\\
41& [42,24]& [43,25]& [44,26]& [45,27]& [46,28]& $\fbox{[47,29]}$& [48,29]& [49,30]& [50,30]& 52A53&No\\
42& [43,25]& [44,26]& [45,27]& [46,28]& [47,29]& [48,29]& [49,30]& [50,31]& $\fbox{[51,32]}$& AA377&Yes\\
43& [44,26]& [45,27]& [46,28]& [47,29]& [48,30]& [49,30]& [50,31]& $\fbox{[51,32]}$& [52,32]& AA377&Yes\\
44& [45,27]& [46,28]& [47,29]& [48,30]& [49,31]& $\fbox{[50,32]}$& [51,32]& [52,32]& [53,33]& 5CBFF&No\\
45& [46,28]& [47,29]& [48,30]& [49,31]& [50,32]& $\fbox{[51,33]}$& [52,33]& [53,34]& [54,34]& 713CF&No\\
46& [47,29]& [48,30]& [49,31]& [50,32]& [51,33]& [52,33]& [53,34]& [54,35]& $\fbox{[55,36]}$& A7DB5&No\\
47& [48,30]& [49,31]& [50,32]& [51,33]& [52,34]& [53,34]& [54,35]& $\fbox{[55,36]}$& [56,36]& A7DB5&No\\
48& [49,31]& [50,32]& [51,33]& [52,34]& [53,34]& [54,35]& $\fbox{[55,36]}$& [56,36]& [57,37]& 80203&No\\
49& [50,32]& [51,33]& [52,34]& [53,35]& $\fbox{[54,36]}$& [55,36]& [56,37]& [57,37]& [58,38]& 5CBFF&No\\
50& [51,33]& [52,34]& [53,35]& [54,36]& $\fbox{[55,37]}$& [56,37]& [57,38]& [58,38]& [59,39]& 5CBFF&No\\
51& [52,34]& [53,35]& [54,36]& $\fbox{[55,37]}$& [56,37]& [57,38]& [58,39]& [59,39]& [60,39]& 5CBFF&No\\
52& [53,35]& [54,36]& [55,37]& [56,37]& [57,38]& [58,39]& $\fbox{[59,40]}$& [60,40]& [61,41]& 82ECB&No\\
53& [54,36]& [55,37]& [56,37]& [57,38]& [58,39]& [59,40]& [60,41]& [61,41]& $\fbox{[62,43]}$& 16758F&No\\
54& [55,37]& [56,38]& [57,38]& [58,39]& [59,40]& [60,41]& [61,42]& [62,43]& $\fbox{[63,44]}$& 804EB&Yes\\
55& [56,38]& [57,38]& [58,39]& [59,40]& [60,41]& [61,42]& [62,43]& $\fbox{[63,44]}$& [64,44]& 804EB&Yes\\
56& [57,38]& [58,39]& [59,40]& [60,41]& [61,42]& [62,43]& [63,44]& $\fbox{[64,45]}$& [65,45]& DC37F&No\\
57& [58,39]& [59,40]& [60,41]& [61,42]& [62,43]& [63,44]& $\fbox{[64,45]}$& [65,45]& [66,45]& 8AE17&No\\
58& [59,40]& [60,41]& [61,42]& [62,43]& [63,44]& [64,45]& $\fbox{[65,46]}$& [66,46]& [67,46]& 8DA3B&No\\
59& [60,41]& [61,42]& [62,43]& [63,44]& [64,45]& [65,46]& $\fbox{[66,47]}$& [67,47]& [68,47]& 864ED&No\\
60& [61,42]& [62,43]& [63,44]& [64,45]& [65,46]& [66,47]& $\fbox{[67,48]}$& [68,48]& [69,48]& 89E97&No\\
\hline
\end{tabular}
}
\end{center}
\begin{center}
\caption{Optimal (shortened) cyclic codes correcting bursts of length up
to 9, for a guard\newline space from $g = 20$ to $g = 60$}
\end{center}
\end{table}

\begin{table}
\begin{center}
\tiny{
\begin{tabular}{|c|c|c|c|c|c|c|c|c|c|c|c|}
\hline
$g$ &$9_1$ &$9_2$ &$9_3$ &$9_4$ &$9_5$ &$9_6$ &$9_7$ &$9_8$ &$9_9$&Polynomial &Cyclical?\\
\hline
61& [62,43]& [63,44]& [64,45]& [65,46]& [66,47]& [67,48]& $\fbox{[68,49]}$& [69,49]& [70,49]& 8420B&No\\
62& [63,44]& [64,45]& [65,46]& [66,47]& [67,48]& [68,49]& $\fbox{[69,50]}$& [70,51]& [71,51]& 9F7D5&No\\
63& [64,45]& [65,46]& [66,47]& [67,48]& [68,49]& [69,50]& $\fbox{[70,51]}$& [71,51]& [72,52]& E62AF&No\\
64& [65,46]& [66,47]& [67,48]& [68,49]& [69,50]& [70,51]& $\fbox{[71,52]}$& [72,52]& [73,52]& 81585&No\\
65& [66,47]& [67,48]& [68,49]& [69,50]& [70,51]& $\fbox{[71,52]}$& [72,52]& [73,53]& [74,54]& 81585&No\\
66& [67,48]& [68,49]& [69,50]& [70,51]& [71,52]& $\fbox{[72,53]}$& [73,53]& [74,54]& [75,54]& 8A785&No\\
67& [68,49]& [69,50]& [70,51]& [71,52]& [72,53]& $\fbox{[73,54]}$& [74,54]& [75,55]& [76,55]& 822B9&No\\
68& [69,50]& [70,51]& [71,52]& [72,53]& [73,54]& [74,55]& $\fbox{[75,56]}$& [76,56]& [77,56]& C634F&No\\
69& [70,51]& [71,52]& [72,53]& [73,54]& [74,55]& [75,56]& $\fbox{[76,57]}$& [77,57]& [78,57]& C3A2F&No\\
70& [71,52]& [72,53]& [73,54]& [74,55]& [75,56]& [76,57]& $\fbox{[77,58]}$& [78,58]& [79,58]& 9860F&No\\
71& [72,53]& [73,54]& [74,55]& [75,56]& [76,57]& [77,58]& $\fbox{[78,59]}$& [79,59]& [80,59]& A7D55&No\\
72& [73,54]& [74,55]& [75,56]& [76,57]& [77,58]& [78,59]& $\fbox{[79,60]}$& [80,60]& [81,60]& 9E245&No\\
73& [74,55]& [75,56]& [76,57]& [77,58]& [78,59]& $\fbox{[79,60]}$& [80,60]& [81,61]& [82,61]& 815E3&No\\
74& [75,56]& [76,57]& [77,58]& [78,59]& [79,60]& [80,61]& $\fbox{[81,62]}$& [82,62]& [83,62]& 84BCB&No\\
75& [76,57]& [77,58]& [78,59]& $\fbox{[79,60]}$& [80,61]& [81,62]& [82,62]& [83,63]& [84,63]& 815E3&No\\
76& [77,58]& [78,59]& [79,60]& [80,61]& [81,62]& $\fbox{[82,63]}$& [83,63]& [84,64]& [85,65]& 887A5&No\\
77& [78,59]& [79,60]& [80,61]& [81,62]& [82,63]& $\fbox{[83,64]}$& [84,64]& [85,65]& [86,65]& 8A3F1&No\\
78& [79,60]& [80,61]& [81,62]& [82,63]& [83,64]& $\fbox{[84,65]}$& [85,65]& [86,66]& [87,66]& 924FD&No\\
79& [80,61]& [81,62]& [82,63]& [83,64]& [84,65]& $\fbox{[85,66]}$& [86,66]& [87,67]& [88,67]& 9C3B5&No\\
80& [81,62]& [82,63]& [83,64]& [84,65]& [85,66]& $\fbox{[86,67]}$& [87,67]& [88,68]& [89,68]& 864ED&No\\
81& [82,63]& [83,64]& [84,65]& [85,66]& [86,67]& [87,68]& $\fbox{[88,69]}$& [89,69]& [90,70]& EAB9F&No\\
82& [83,64]& [84,65]& [85,66]& $\fbox{[86,67]}$& [87,68]& [88,69]& [89,69]& [90,70]& [91,70]& 80C0B&No\\
83& [84,65]& [85,66]& [86,67]& [87,68]& [88,69]& $\fbox{[89,70]}$& [90,70]& [91,71]& [92,71]& FAE3F&No\\
84& [85,66]& [86,67]& [87,68]& [88,69]& [89,70]& $\fbox{[90,71]}$& [91,71]& [92,72]& [93,73]& FAE3F&No\\
85& [86,67]& [87,68]& [88,69]& [89,70]& [90,71]& $\fbox{[91,72]}$& [92,72]& [93,73]& [94,73]& 924FD&No\\
86& [87,68]& [88,69]& [89,70]& [90,71]& $\fbox{[91,72]}$& [92,72]& [93,73]& [94,74]& [95,74]& 8420B&No\\
87& [88,69]& [89,70]& [90,71]& [91,72]& [92,73]& $\fbox{[93,74]}$& [94,74]& [95,75]& [96,75]& 8A3F1&No\\
88& [89,70]& [90,71]& [91,72]& [92,73]& [93,74]& $\fbox{[94,75]}$& [95,75]& [96,76]& [97,76]& 907CF&No\\
89& [90,71]& [91,72]& [92,73]& [93,74]& $\fbox{[94,75]}$& [95,76]& [96,76]& [97,76]& [98,77]& 8BE39&No\\
90& [91,72]& [92,73]& [93,74]& [94,75]& $\fbox{[95,76]}$& [96,76]& [97,77]& [98,77]& [99,78]& 8A3F1&No\\
91& [92,73]& [93,74]& [94,75]& [95,76]& [96,77]& $\fbox{[97,78]}$& [98,78]& [99,78]& [100,79]& CD7D8&No\\
92& [93,74]& [94,75]& [95,76]& [96,77]& $\fbox{[97,78]}$& [98,78]& [99,79]& [100,79]& [101,80]& 8AA4F&No\\
93& [94,75]& [95,76]& [96,77]& [97,78]& [98,79]& $\fbox{[99,80]}$& [100,80]& [101,80]& [102,81]& 8BE39&No\\
94& [95,76]& [96,77]& [97,78]& [98,79]& $\fbox{[99,80]}$& [100,80]& [101,81]& [102,82]& [103,82]& 8BE39&No\\
95& [96,77]& [97,78]& [98,79]& [99,80]& $\fbox{[100,81]}$& [101,81]& [102,82]& [103,82]& [104,83]& 907CF&No\\
96& [97,78]& [98,79]& [99,80]& [100,81]& [101,82]& [102,83]& [103,84]& [104,85]& $\fbox{[105,86]}$& CA2CB&Yes\\
97& [98,79]& [99,80]& [100,81]& [101,82]& [102,83]& [103,84]& [104,85]& $\fbox{[105,86]}$& [106,86]& CA2CB&Yes\\
98& [99,80]& [100,81]& [101,82]& [102,83]& [103,84]& [104,85]& $\fbox{[105,86]}$& [106,86]& [107,86]& CA2CB&Yes\\
99& [100,81]& [101,82]& [102,83]& [103,84]& [104,85]& $\fbox{[105,86]}$& [106,86]& [107,86]& [108,87]& CA2CB&Yes\\
100& [101,82]& [102,83]& [103,84]& [104,85]& $\fbox{[105,86]}$& [106,86]& [107,87]& [108,87]& [109,87]& CA2CB&Yes\\
\hline
\end{tabular}
}
\end{center}
\begin{center}
\caption{Optimal (shortened) cyclic codes correcting bursts of length up
to 9, for a guard\newline space from $g = 64$ to $g = 100$}
\end{center}
\end{table}

\begin{table}
\begin{center}
\tiny{
\begin{tabular}{|c|c|c|c|c|c|c|c|c|c|c|c|c|}
\hline
$g$ &$10_1$ &$10_2$ &$10_3$ &$10_4$ &$10_5$ &$10_6$ &$10_7$ &$10_8$ &$10_9$&$10_{10}$&Polynomial&Cyclical?\\
\hline
20&[21,1]&[22,2]&[23,3]&[24,4]&[25,5]&[26,6]&[27,7]&[28,8]&[29,9]&\fbox{[30,10]}&100401&Yes\\
21&[22,2]&[23,3]&[24,4]&[25,5]&[26,6]&[27,7]&[28,8]&[29,9]&[30,10]&\fbox{[31,11]}&11A799&Yes\\
22&[23,3]&[24,4]&[25,5]&[26,6]&[27,7]&[28,8]&[29,9]&[30,10]&[31,11]&\fbox{[32,12]}&118C05&No\\
23&[24,4]&[25,5]&[26,6]&[27,7]&[28,8]&[29,9]&[30,10]&[31,11]&[32,12]&\fbox{[33,12]}&20AB87&No\\
24&[25,5]&[26,6]&[27,7]&[28,8]&[29,9]&[30,10]&[31,11]&[32,12]&\fbox{[33,13]}&[34,13]&1154A5&No\\
25&[26,6]&[27,7]&[28,8]&[29,9]&[30,10]&[31,11]&[32,12]&[33,13]&[34,14]&\fbox{[35,15]}&100421&Yes\\
26&[27,7]&[28,8]&[29,9]&[30,10]&[31,11]&[32,12]&[33,13]&[34,14]&\fbox{[35,15]}&[36,15]&2087E3&No\\
27&[28,8]&[29,9]&[30,10]&[31,11]&[32,12]&[33,13]&[34,14]&[35,15]&\fbox{[36,16]}&[37,16]&266E55&No\\
28&[29, 9]&[30, 10]&[31, 11]&[32, 12]&[33, 13]&[34, 14]&[35, 15]&[36, 16]&\fbox{[37, 17]}&[38, 17]&100439&No\\
29&[30, 10]&[31, 11]&[32, 12]&[33, 13]&[34, 14]&[35, 15]&[36, 16]&\fbox{[37, 17]}&[38, 18]&[39, 17]&100429&No\\
30&[31, 11]&[32, 12]&[33, 13]&[34, 14]&[35, 15]&[36, 16]&[37, 17]&\fbox{[38, 18]}&[39, 18]&[40, 18]&10043F&No\\
31&[32, 12]&[33, 13]&[34, 14]&[35, 15]&[36, 16]&[37, 17]&[38, 18]&\fbox{[39, 19]}&[40, 19]& [41,19] &10043F&No\\
32&[33,13] & [34,14] & [35,15] & [36,16] & [37,17] & [38,18] & [39,19] & [40,20] & [41,20] &\fbox{ [42,21] }&200401&No\\
33&[34, 14]&[35, 15]&[36, 16]&[37, 17]&[38, 18]&[39, 19]&[40, 20]&\fbox{[41, 21]}&[42, 21]&[43, 22]&1387AF&No\\
34&[35, 15]&[36, 16]&[37, 17]&[38, 18]&[39, 19]&[40, 20]&[41, 21]&\fbox{[42, 22]}&[43, 22]&[44, 23]&162C1B&No\\
35&[36, 16]&[37, 17]&[38, 18]&[39, 19]&[40, 20]&[41, 21]&[42, 22]&\fbox{[43, 23]}&[44, 23]&[45, 24]&100563&No\\
36&[37, 17]&[38, 18]&[39, 19]&[40, 20]&[41, 21]&[42, 22]&[43, 23]&\fbox{[44, 24]}&[45, 24]&[46, 24]&120447&No\\
37&[38, 18]&[39, 19]&[40, 20]&[41, 21]&[42, 22]&[43, 23]&[44, 24]&\fbox{[45, 25]}&[46, 25]&[47, 25]&16E7CF&No\\
38&[39, 19]&[40, 20]&[41, 21]&[42, 22]&[43, 23]&[44, 24]&[45, 25]&\fbox{[46, 26]}&[47, 26]&[48, 26]&108411&No\\
39&[40, 20]&[41, 21]&[42, 22]&[43, 23]&[44, 24]&[45, 25]&\fbox{[46, 26]}&[47, 26]&[48, 27]&[49, 27]&108411&No\\
40&[41, 21]&[42, 22]&[43, 23]&[44, 24]&[45, 25]&[46, 26]&[47, 27]&\fbox{[48, 28]}&[49, 28]&[50, 28]&147DEF&No\\
41&[42,22]&[43,23]&[44,24]&[45,25]&[46,26]&[47,27]&[48,28]&[49,28]&[50,29]&\fbox{[51,30]}&200C03&No\\
42&[43, 23]&[44, 24]&[45, 25]&[46, 26]&[47, 27]&[48, 28]&\fbox{[49, 29]}&[50, 29]&[51, 30]&[52, 30]&18064F&No\\
43&[44, 24]&[45, 25]&[46, 26]&[47, 27]&[48, 28]&[49, 29]&[50, 30]&[51, 30]&\fbox{[52, 31]}&[53, 31]&20A5D5&No\\
44&[45, 25]&[46, 26]&[47, 27]&[48, 28]&[49, 29]&[50, 30]&[51, 30]&[52, 31]&\fbox{[53, 32]}&[54, 32]&29FA3D&No\\
45&[46, 26]&[47, 27]&[48, 28]&[49, 29]&[50, 30]&[51, 31]&[52, 31]&[53, 32]&[54, 33]&\fbox{[55, 34]}&253BB7&Yes\\
46&[47, 27]&[48, 28]&[49, 29]&[50, 30]&[51, 31]&[52, 32]&[53, 32]&[54, 33]&\fbox{[55, 34]}&[56, 34]&234F7D&No\\
47&[48, 28]&[49, 29]&[50, 30]&[51, 31]&[52, 32]&[53, 33]&[54, 33]&[55, 34]&\fbox{[56, 35]}&[57, 35]&34F557&No\\
48&[49, 29]&[50, 30]&[51, 31]&[52, 32]&[53, 33]&[54, 34]&[55, 34]&[56, 35]&\fbox{[57, 36]}&[58, 36]&3A874F&No\\
49&[50, 30]&[51, 31]&[52, 32]&[53, 33]&[54, 34]&[55, 34]&[56, 35]&[57, 36]&\fbox{[58, 37]}&[59, 37]&3A874F&No\\
50&[51, 31]&[52, 32]&[53, 33]&[54, 34]&[55, 35]&\fbox{[56, 36]}&[57, 36]&[58, 37]&[59, 37]&[60, 38]&14B7D5&No\\
51&[52, 32]&[53, 33]&[54, 34]&[55, 35]&[56, 36]&[57, 37]&[58, 37]&\fbox{[59, 38]}&[60, 39]&[61, 38]&204427&No\\
52&[53, 33]&[54, 34]&[55, 35]&[56, 36]&[57, 37]&[58, 37]&[59, 38]&[60, 39]&\fbox{[61, 39]}&[62,40]&200E33&No\\
53&[54,34]&[55,35]&[56,36]&[57,37]&[58,38]&[59,38]&[60,39]&[61,40]&[62,41]&\fbox{[63,42]}&22FDB7&Yes\\
54&[55, 35]&[56, 36]&[57, 37]&[58, 38]&[59, 38]&[60, 39]&[61, 40]&[62, 41]&\fbox{[63, 42]}&[64, 42]&22FDB7&Yes\\
55&[56, 36]&[57, 37]&[58, 38]&[59, 38]&[60, 39]&[61, 40]&[62, 41]&\fbox{[63, 42]}&[64, 42]&[65, 42]&8005DF&No\\
56&[57, 37]&[58, 38]&[59, 39]&[60, 39]&[61, 40]&[62, 41]&[63, 42]&\fbox{[64, 43]}&[65, 43]&[66, 43]&20F983&No\\
57&[58, 38]&[59, 39]&[60, 39]&[61, 40]&[62, 41]&[63, 42]&[64, 43]&\fbox{[65, 44]}&[66, 44]&[67, 44]&20042B&No\\
58&[59, 39]&[60, 39]&[61, 40]&[62, 41]&[63, 42]&[64, 43]&[65, 44]&\fbox{[66, 45]}&[67, 45]&[68, 45]&208811&No\\
59&[60, 39]&[61, 40]&[62, 41]&[63, 42]&[64, 43]&[65, 44]&[66, 45]&\fbox{[67, 46]}&[68, 46]&[69, 47]&209DD5&No\\
60&[61, 40]&[62, 41]&[63, 42]&[64, 43]&[65, 44]&[66, 45]&[67, 46]&\fbox{[68, 47]}&[69, 47]&[70, 48]&2114F7&No\\
\hline
\end{tabular}
}
\end{center}
\begin{center}
\caption{Optimal (shortened) cyclic codes correcting bursts of length up
to 10, for a guard\newline space from $g = 20$ to $g = 60$}
\end{center}
\end{table}

\begin{table}
\begin{center}
\tiny{
\begin{tabular}{|c|c|c|c|c|c|c|c|c|c|c|c|c|}
\hline
$g$ &$10_1$ &$10_2$ &$10_3$ &$10_4$ &$10_5$ &$10_6$ &$10_7$ &$10_8$ &$10_9$&$10_{10}$&Polynomial&Cyclical?\\
\hline
61&[62,41]&[63,42]&[64,43]&[65,44]&[66,45]&[67,46]&[68,47]&\fbox{[69,48]}&[70,48]&[71,49]&21EE9F&No\\
62&[63,42]&[64,43]&[65,44]&[66,45]&[67,46]&[68,47]&[69,48]&\fbox{[70,49]}&[71,49]&[72,50]&256F93&No\\
63&[64,43]&[65,44]&[66,45]&[67,46]&[68,47]&[69,48]&[70,49]&\fbox{[71,50]}&[72,50]&[73,51]&22BDB1&No\\
64&[65,44]&[66,45]&[67,46]&[68,47]&[69,48]&[70,49]&[71,50]&\fbox{[72,51]}&[73,51]&[74,52]&209947&No\\
65&[66,45]&[67,46]&[68,47]&[69,48]&[70,49]&[71,50]&[72,51]&\fbox{[73,52]}&[74,52]&[75,53]&3BCB6F&No\\
66&[67,46]&[68,47]&[69,48]&[70,49]&[71,50]&[72,51]&[73,52]&\fbox{[74,53]}&[75,53]&[76,54]&200865&No\\
67&[68,47]&[69,48]&[70,49]&[71,50]&[72,51]&[73,52]&\fbox{[74,53]}&[75,54]&[76,54]&[77,55]&200865&No\\
68&[69,48]&[70,49]&[71,50]&[72,51]&[73,52]&[74,53]&[75,54]&\fbox{[76,55]}&[77,55]&[78,56]&205653&No\\
69&[70,49]&[71,50]&[72,51]&[73,52]&[74,53]&[75,54]&[76,55]&\fbox{[77,56]}&[78,56]&[79,56]&3CA46F&No\\
70&[71,50]&[72,51]&[73,52]&[74,53]&[75,54]&[76,55]&[77,56]&\fbox{[78,57]}&[79,57]&[80,58]&2A2ED3&No\\
71&[72,51]&[73,52]&[74,53]&[75,54]&[76,55]&[77,56]&\fbox{[78,57]}&[79,57]&[80,58]&[81,58]&2006F9&No\\
72&[73,52]&[74,53]&[75,54]&[76,55]&[77,56]&[78,57]&[79,58]&\fbox{[80,59]}&[81,59]&[82,60]&23EF19&No\\
73&[74,53]&[75,54]&[76,55]&[77,56]&[78,57]&[79,58]&\fbox{[80,59]}&[81,59]&[82,60]&[83,60]&213795&No\\
74&[75,54]&[76,55]&[77,56]&[78,57]&[79,58]&[80,59]&\fbox{[81,60]}&[82,60]&[83,61]&[84,62]&209DD5&No\\
75&[76,55]&[77,56]&[78,57]&[79,58]&[80,59]&[81,60]&\fbox{[82,61]}&[83,62]&[84,62]&[85,63]&200D25&No\\
76&[77,56]&[78,57]&[79,58]&[80,59]&[81,60]&[82,61]&[83,62]&[84,63]&\fbox{[85,64]}&[86,64]&2F571B&No\\
77&[78,57]&[79,58]&[80,59]&[81,60]&[82,61]&[83,62]&[84,63]&\fbox{[85,64]}&[86,64]&[87,65]&2F16E7&No\\
78&[79,58]&[80,59]&[81,60]&[82,61]&[83,62]&[84,63]&[85,64]&\fbox{[86,65]}&[87,65]&[88,66]&2006F9&No\\
79&[80,59]&[81,60]&[82,61]&[83,62]&[84,63]&[85,64]&\fbox{[86,65]}&[87,65]&[88,66]&[89,67]&2006F9&No\\
80&[81,60]&[82,61]&[83,62]&[84,63]&[85,64]&[86,65]&\fbox{[87,66]}&[88,66]&[89,67]&[90,67]&2148F5&No\\
81&[82,61]&[83,62]&[84,63]&[85,64]&[86,65]&[87,66]&\fbox{[88,67]}&[89,67]&[90,68]&[91,69]&2006F9&No\\
82&[83,62]&[84,63]&[85,64]&[86,65]&[87,66]&[88,67]&\fbox{[89,68]}&[90,68]&[91,69]&[92,70]&25AFFD&No\\
83&[84,63]&[85,64]&[86,65]&[87,66]&[88,67]&[89,68]&[90,69]&[91,70]&[92,71]&\fbox{[93,72]}&296957&Yes\\
84&[85,64]&[86,65]&[87,66]&[88,67]&[89,68]&[90,69]&[91,70]&[92,71]&\fbox{[93,72]}&[94,72]&296957&Yes\\
85&[86,65]&[87,66]&[88,67]&[89,68]&[90,69]&[91,70]&[92,71]&\fbox{[93,72]}&[94,72]&[95,73]&296957&Yes\\
86&[87,66]&[88,67]&[89,68]&[90,69]&[91,70]&[92,71]&\fbox{[93,72]}&[94,72]&[95,73]&[96,73]&27FB4B&No\\
87&[88,67]&[89,68]&[90,69]&[91,70]&[92,71]&[93,72]&\fbox{[94,73]}&[95,73]&[96,74]&[97,74]&213995&No\\
88&[89,68]&[90,69]&[91,70]&[92,71]&[93,72]&[94,73]&\fbox{[95,74]}&[96,74]&[97,75]&[98,75]&2E0B93&No\\
89&[90,69]&[91,70]&[92,71]&[93,72]&[94,73]&\fbox{[95,74]}&[96,74]&[97,75]&[98,76]&[99,77]&4004D5&No\\
90&[91,70]&[92,71]&[93,72]&[94,73]&[95,74]&[96,75]&\fbox{[97,76]}&[98,76]&[99,77]&[100,78]&336EAF&No\\
91&[92,71]&[93,72]&[94,73]&[95,74]&[96,75]&[97,76]&\fbox{[98,77]}&[99,77]&[100,78]&[101,79]&2896CD&No\\
92&[93,72]&[94,73]&[95,74]&[96,75]&[97,76]&\fbox{[98,77]}&[99,77]&[100,78]&[101,79]&[102,80]&20A9C7&No\\
93&[94,73]&[95,74]&[96,75]&[97,76]&[98,77]&[99,78]&\fbox{[100,79]}&[101,79]&[102,80]&[103,81]&400447&No\\
94&[95,74]&[96,75]&[97,76]&[98,77]&[99,78]&[100,79]&\fbox{[101,80]}&[102,80]&[103,81]&[104,82]&330FDF&No\\
95&[96,75]&[97,76]&[98,77]&[99,78]&[100,79]&[101,80]&[102,81]&[103,82]&[104,83]&\fbox{[105,84]}&330FDF&Yes\\
96&[97,76]&[98,77]&[99,78]&[100,79]&[101,80]&[102,81]&[103,82]&[104,83]&\fbox{[105,84]}&[106,84]&330FDF&Yes\\
97&[98,77]&[99,78]&[100,79]&[101,80]&[102,81]&[103,82]&[104,83]&\fbox{[105,84]}&[106,83]&[107,84]&330FDF&Yes\\
98&[99,78]&[100,79]&[101,80]&[102,81]&[103,82]&[104,83]&[105,84]&\fbox{[106,85]}&[107,84]&[108,85]&20A9C7&No\\
99&[100,79]&[101,80]&[102,81]&[103,82]&[104,83]&[105,84]&\fbox{[106,85]}&[107,85]&[108,85]&[109,86]&20A9C7&No\\
100&[101,80]&[102,81]&[103,82]&[104,83]&[105,84]&\fbox{[106,85]}&[107,85]&[108,86]&[109,86]&[110,86]&20A9C7&No\\
\hline
\end{tabular}
}
\end{center}
\begin{center}
\caption{Optimal (shortened) cyclic codes correcting bursts of length up
to 10, for a guard\newline space from $g = 61$ to $g = 100$}
\end{center}
\end{table}


\vspace{.5cm}


\section{Conclusions}
We have presented an efficient algorithm finding the best cyclic or
shortened cyclic
single burst-correcting codes  for different parameters. The algorithm
minimizes the number of syndrome checks by using Gray codes.
Extensive tables with the most efficient codes have been presented.

\end{document}